\documentclass[11pt]{article}
\usepackage{a4wide}
\usepackage{amsmath,amssymb,amsfonts}
\usepackage{subcaption}
\usepackage{comment}
\usepackage{epsfig,verbatim,graphics}
\usepackage{slashed,siunitx}
\usepackage[table,x11names]{xcolor}
\usepackage{wrapfig}

\usepackage{tikz} 
\usepackage{tikz-feynman}
\tikzfeynmanset{compat=1.0.0}
\usetikzlibrary{shapes,arrows,positioning,automata,backgrounds,calc,er,patterns}
\newcommand{\mathsym}[1]{{}}

\usepackage[utf8]{luainputenc}
\usepackage{esint}

\newcommand{\eref}[1]{(\ref{#1})}
\renewcommand\({\left(}
\renewcommand\){\right)}
\renewcommand\[{\left[}
\renewcommand\]{\right]}

\newcommand{\dd}{{\rm d}}
\declareslashed{}{{^-}}{.1}{0.1}{\rm d}

\newcommand\vp{\varphi}
\newcommand\eps{\epsilon}
\newcommand\mpl{m_{\rm P}}

\def\ba{\begin{eqnarray}}
\def\ea{\end{eqnarray}}
\def\be{\begin{equation}}
\def\ee{\end{equation}}

\def\L{\mathcal{L}}
\def\O{\mathcal{O}}

\def\M{\mathcal{M}}

\def\G{\mathcal{G}}

\def\del{\nabla}
\def\nn{\nonumber}
\def\({\left(}
\def\){\right)}

\def\eref#1{(\ref{#1})}

\usepackage{tikz}
\usetikzlibrary{arrows,shapes}
\usetikzlibrary{trees}
\usetikzlibrary{matrix,arrows} 				
\usetikzlibrary{positioning}				
\usetikzlibrary{calc,through}				
\usetikzlibrary{decorations.pathreplacing}  
\usepackage{pgffor}							
\usetikzlibrary{decorations.pathmorphing}	
\usetikzlibrary{decorations.markings}
\tikzset{
    vector/.style={decorate, decoration={snake}, draw},
	provector/.style={decorate, decoration={snake,amplitude=2.5pt}, draw},
	antivector/.style={decorate, decoration={snake,amplitude=-2.5pt}, draw},
    fermion/.style={draw=black, postaction={decorate},
        decoration={markings,mark=at position .55 with {\arrow[draw=black]{>}}}},
    fermionbar/.style={draw=black, postaction={decorate},
        decoration={markings,mark=at position .55 with {\arrow[draw=black]{<}}}},
    fermionnoarrow/.style={draw=black},
    gluon/.style={decorate, draw=black,
        decoration={coil,amplitude=4pt, segment length=5pt}},
    scalar/.style={dashed,draw=black, postaction={decorate},
        decoration={markings,mark=at position .55 with {\arrow[draw=black]{>}}}},
    scalarbar/.style={dashed,draw=black, postaction={decorate},
        decoration={markings,mark=at position .55 with {\arrow[draw=black]{<}}}},
    scalarnoarrow/.style={dashed,draw=black},
    electron/.style={draw=black, postaction={decorate},
        decoration={markings,mark=at position .55 with {\arrow[draw=black]{>}}}},
	bigvector/.style={decorate, decoration={snake,amplitude=4pt}, draw},
}
\newcommand{\roughly}[1]{\mathrel{\raise.3ex\hbox{$#1$\kern-0.85em
\lower1ex\hbox{$\sim$}}}}
\usepackage{jheppub}
\usepackage{cleveref}
\begin{document}
\begin{titlepage}
\begin{center}

\hfill Nikhef 2020-013

\vskip 2cm

{\huge \bf  
  Matching and running sensitivity\\[2mm] in non-renormalizable inflationary models        
}
\vskip 1cm

\renewcommand*{\thefootnote}{\fnsymbol{footnote}}
\setcounter{footnote}{0}

{\bf\large{
	Jacopo\ Fumagalli$^{1}$\footnote{{\tt fumagall@iap.fr }
	        \footnotemark[2]{\tt mpostma@nikhef.nl},
	        \footnotemark[3]{\tt melvinvandenbout@gmail.com}},
	Marieke Postma$^{2}$\footnotemark[2] 
	and
	Melvin van den Bout$^2$\footnotemark[3] 
}}

\renewcommand*{\thefootnote}{\number{footnote}}
\setcounter{footnote}{0}

\vskip 25pt

{\em $^1$ \hskip -.1truecm
	Institut d'Astrophysique de Paris, GReCO, UMR 7095 du CNRS et de Sorbonne Universit\'e, 98bis boulevard Arago, 75014 Paris, France
}

\vskip 20pt
{
\em $^2$ \hskip -.1truecm
Nikhef, \\Science Park 105, \\1098 XG Amsterdam, The Netherlands}

\vskip 1cm

{\bf ABSTRACT}\\[3ex]
\end{center} Most of the inflationary models that are in agreement with the Planck data rely on the presence of non-renormalizable operators. If the connection to low energy particle physics is made, the renormalization group (RG) introduces a sensitivity to ultraviolet (UV) physics that can be crucial in determining the inflationary predictions. We analyse this effect for the Standard Model (SM) augmented with non-minimal derivative couplings to gravity.  Our set-up reduces to the SM for small values of the Higgs field, and allows for inflation in the opposite large field regime.  The one-loop beta functions in the inflationary region are calculated using a covariant approach that properly accounts for the non-trivial structure of the field space manifold.  We run the SM parameters from the electroweak to the inflationary scale, matching the couplings of the different effective field theories at the boundary between the two regimes, where we also include threshold corrections that parametrize effects from UV physics. We then compute the spectral index and tensor-to-scalar ratio and find that RG flow corrections can be determinant: a scenario that is ruled out at tree level can be resurrected and vice versa.  \end{titlepage}

\newpage
\setcounter{page}{1} \tableofcontents

\newpage

\section{Introduction}

One of the main lessons learned from the Planck constraints on inflation \cite{Akrami:2018odb} is that quadratic and quartic inflation, arguably the simplest approaches, are ruled out by the data. Most of the successful inflationary models, whether single or multi-field, instead rely on non-renormalizable operators to obtain predictions in agreement with Planck (see \cite{Martin:2013tda} for a large list of models).

It is well known that inflation is sensitive to ultraviolet (UV) physics. Corrections from high energy degrees of freedom tend to increase the inflaton mass, thus ruining the inflationary dynamics needed to sustain a long enough period of inflation \cite{Copeland:1994vg}.  There are two aspects to this so-called eta-problem (see \cite{Baumann:2014nda} for a review). First, integrating out heavy physics above a given cutoff scale shifts the parameters in the low energy effective field theory (EFT) by an amount proportional to the strong coupling scale. That is the famous hierarchy problem in the context of cosmology. Second, even Planck suppressed irrelevant operators can easily spoil the flatness of the inflaton potential and completely change the inflationary dynamics.

In this work we focus on a different kind of UV sensitivity which stems from the non-renormalizable character of an inflationary model and comes into play when the connection to the low-energy (beyond the) Standard Model (SM) degrees of freedom is made using the renormalization group (RG) flow \cite{cliffnew,Fumagalli}. When the RG improved action is used to incorporate perturbative quantum corrections, the running of the couplings can affect the naive tree level inflationary predictions. To compute meaningful observables one must determine the RG equations in the inflationary regime, and understand the effects of UV physics at the cutoff scale ${\cal M}$, set by the non-renormalizable operators, on the running.  Although our approach to analyse this effect is generic, for concreteness we will concentrate on the SM non-minimally coupled to gravity via derivative interactions \cite{Germani:2010gm,Germani:2011cv,disformal}.

The presence of a non-renormalizable operators allows to distinguish different regimes characterized by small (field-dependent) parameters. To be concrete, consider the toy model Lagrangian
\be
\mathcal{L}=-\frac{1}{2}\(1+\frac{\phi}{{\cal M}}\)^2(\partial\phi)^2+V(\phi).
\label{L_toy}
\ee
In the small field regime $\delta \equiv \phi/\mathcal{M} \ll 1$ the higher-dimensional interaction is a small correction, whereas in the large field regime $\delta^{-1} \ll 1$ it gives the dominant contribution to the kinetic term. In both domains the model may be \textit{renormalizable in the EFT sense}, by which we mean that in each field region\footnote{If there are several non-renormalizable operators there can be more than two regimes. Important is that the small field and the inflationary regime both have renormalizable EFT descriptions.  On the boundaries, and in the midfield ranges where the EFT description fails, UV physics becomes relevant.} it is possible to define a small parameter in such a way that at every order in that parameter the theory can be renormalized with a finite number of counterterms. In the example above, loop corrections can be organized in a series expansion in $\delta$ ($\delta^{-1}$) in the small (large) field regime.  This can be seen, for instance, by considering the one-loop contribution to the effective potential
\be
V^{\mathrm{1-loop}}= \frac{(V'')^2}{32\pi^2\epsilon}=\frac{2\lambda V_{\mathrm{tree}}}{\pi^2\epsilon}\frac{\(\delta+\frac{3}{2}\)^2}{(\delta+1)^6}
=\frac{\lambda V_{\mathrm{tree}}}{\pi^2\epsilon}\cdot
\left\{ \begin{array}{ll}
	 \sum_{k=0}^{\infty}c_k\delta^k&\quad\delta\equiv \phi/\Lambda \ll 1 \\
	\sum_{k=0}^{\infty}\bar{c}_k(\delta^{-1})^k, &\quad \delta^{-1} \ll 1,
\end{array}\right.                                                                      
\label{toy1}
\ee
where we used dimensional regularization with $d=4-\epsilon$, and a prime denotes a derivative with respect to the canonically normalized field.  However, crossing the boundary between the two regimes at $\delta \approx 1$ a full tower of higher-order operators becomes relevant and the model is not renormalizable in the EFT sense.  It follows that UV physics is important around the boundary and may affect the running of the couplings in this regime appreciably.\footnote{We assume the UV physics does not alter the inflaton potential at tree level, as this would destroy all predictiveness of the model.} This will then modify the value of the parameters in the RG improved inflationary potential, and thus potentially the predictions of the model.

The possibility that inflation can be sensitive to the six-dimensional SM-EFT operators \cite{trott} was first noted in \cite{cliffnew} and further studied in \cite{Fumagalli}. The net effect of those operators is to smear the low energy parameters at the scale marking the boundary between the two regimes. This can equivalently be parametrized in a simpler way by a shift of the low energy parameters at the scale $\mathcal{M}$ \cite{shap,critical1,Enckell:2016xse,Bezrukov:2017dyv} (see also \cite{Fumagalli:2017cdo} for a discussion in the context of new Higgs inflation), an approach we will follow in this paper.

The RG flow analysis is particularly relevant for inflationary models whose parameters are (or can be) measured in low energy experiments such as the Large Hadron Collider.  Prime examples are models where the Higgs boson plays the role of the inflaton \cite{bezrukov1,Germani:2010gm,Kamada:2012se} (see \cite{Rubio:2018ogq} for a recent review); due to an additional non-renormalizable coupling between the SM Higgs boson and the gravity sector, the running of the SM parameters is sensitive to UV physics in the mid-field regime \cite{bezrukov_loop,bezrukov4, wilczek, barvinsky, barvinsky2,barvinsky3,damien,damien2}. The feasibility of connecting low energy physics to inflation was questioned in \cite{Hertzberg2,mirage}, and it has motivated various analyses of UV corrections to the inflationary predictions \cite{cliffnew,Fumagalli,Enckell:2016xse,Fumagalli:2016sof,Bezrukov:2017dyv,Fumagalli:2017cdo}. An interesting approach is also the alternative Palatini formulation of gravity \cite{Bauer:2008zj,Rasanen:2017ivk,Enckell:2018kkc,Rasanen:2018ihz,Racioppi:2019jsp} which gives a higher cutoff scale \cite{Bauer:2010jg} and therefore less sensitivity to UV physics.  It is important to note that the connection to low scale observables is not only an issue for models that embed inflation in (extensions of) the SM; for succesful reheating any inflationary model must be coupled in some way to the SM degrees of freedom.

In this work we continue our investigation of the effects of quantum corrections entering through the RG flow on the inflationary observables pursued in \cite{Fumagalli,Fumagalli:2016sof,Fumagalli:2017cdo}.\footnote{Our study in \cite{Fumagalli} was later implemented (and confirmed where the analyses overlap) by other groups \cite{Enckell:2016xse,Bezrukov:2017dyv,Rasanen:2017ivk,Enckell:2018kkc}. In particular, the authors in \cite{Enckell:2016xse,Bezrukov:2017dyv} study the critical regime where inflation takes place near an inflection point, while \cite{Rasanen:2017ivk,Enckell:2018kkc} analyses the loop corrections in the hilltop scenario in the Palatini formulation.} In particular, we complete the analysis for the SM model non-minimally derivatively coupled to gravity, which goes under the name of new Higgs inflation \cite{Germani:2010gm,Germani:2011cv,disformal}. In previous work \cite{Fumagalli:2017cdo} we showed that the tree level cutoff of the theory is always below the typical energy scales involved at every stage in the universe's history, complementing the analysis in \cite{germaniU}. Furthermore, we pointed out the RG sensitivity of these type of models through analytical considerations (revisited here in \cref{s:flat}).

We will derive the Renormalization group equations (RGEs) at leading order in the large field regime, and we explicitly show under which conditions the spectral index and tensor-to-scalar ratio $(n_s,r)$ are sensitive to the running of the couplings.  Our main results are \cref{inflationRGE1} and figs. \ref{planck1}, \ref{planck2} and \ref{caseb}.  The punchline is that computing predictions at tree level is often not enough. Boundary conditions at the electroweak scale, the unknown UV completion as well as the explicit form of the RGEs could easily have an impact on the inflationary parameters.  This is qualitatively different from plateau-like models of inflation such as Higgs inflation \cite{Fumagalli} and the larger class of Cosmological Attractors \cite{Galante:2014ifa,Fumagalli:2016sof}, where the inflationary predictions are insensitive to running effects to lowest order in slow-roll parameters.

\subsection*{Roadmap}

\begin{figure}[t]
	\centering
	\includegraphics[width=0.80\textwidth]{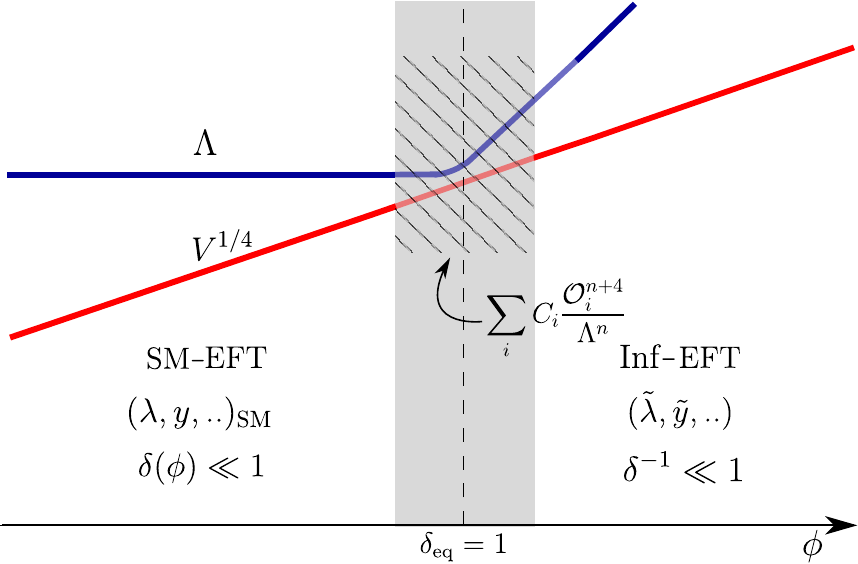} \hspace{0.5cm}
	\caption{
          \textit{
             Overview of the renormalization group flow effects on the inflationary predictions. Shown is the typical energy of the potential $V^{1/4}$ (red) and the unitarity cutoff (blue) as a function of the parameter  $\delta = V/{\cal M}^4$ that distinguishes between the small field ($\delta \ll 1$) and large field ($\delta \gg 1$) region.  In each regime an EFT can be constructed which is valid below the cutoff scale, and which depends on a set of independent couplings that are matched at the boundary at $\delta=1$. In the mid field region, indicated by the grey area, the EFT description breaks down, and the effects of UV physics (parameterized by a tower of higher order operators suppressed by the cutoff scale) can be considerable. We implement this by adding threshold corrections to the couplings at the matching scale.  We calculalate the inflationary RGEs to subsequently run the couplings from the mid-field to the inflationary scale $\delta_\star$, where we calculate $n_s$ and $r$ and analyse the effects of the threshold corrections. }}\label{scheme}
\end{figure}

The RGEs during inflation and the consequences for the observables are derived in a systematic way. We concentrate on the SM with non-minimal derivative couplings to gravity,  reviewed in \cref{themodel}, but our results can be adapted to other set-ups.  We include the possibility that next to the Higgs field, also the fermions and/or gauge bosons have non-minimal derivative couplings. \Cref{scheme} illustrates our approach.

To start, we split the field domain in two regimes in \cref{largefieldsection}, and define the two EFTs in these asymptotic field regions ordered by a small parameter in \cref{largecounter}. We identify the independent set of couplings at leading order in the EFT expansion, which may be different in the two regimes.  To calculate observables, it is not only important to include the effects of UV physics on the running of the couplings in the boundary region, but is also crucial to understand the relevant couplings in the inflationary regime and how they evolve under the RG flow.

The inflationary action in Higgs inflaton models is non-standard with non-trivial kinetic sectors and higher dimensional interactions. For instance, the derivative couplings considered in this paper generate a non-trivial geometry for the field space manifold. To deal with this complication we use a covariant approach \cite{Vilkovisky:1984st,Gong:2011uw,damien2} to compute the one-loop corrections. The formalism is set up in \cref{covariant fields}, while in \cref{s:renormalizability} we compute the one-loop beta functions.

In \cref{pred} we numerically calculate the quantum corrected inflationary predictions for different boundary conditions at the electroweak scale, and for different threshold corrections parameterizing the effects of UV physics on the running in the mid-field region. Starting at the electroweak scale the SM RGEs are used to run the couplings to the boundary between the small and large field regime. At the boundary we match the SM couplings to the set of independent couplings of the inflationary EFT, where we include possible threshold corrections to capture the effects of UV physics. We then run the couplings to the energy scale of inflation using the RGEs derived. The spectral index $n_s$ and tensor-to-scalar ratio $r$ are then computed from the RG improved action, which gives our final result.

We conclude and provide an outlook in \cref{outlook}. 

      \section{The model: Standard Model with non-minimal derivative couplings}\label{themodel}

 We consider the Standard Model (SM) and the Einstein-Hilbert action augmented with
 non-minimally derivative couplings to gravity for the Higgs, gauge and fermion fields.
The action is 
\be
S= \int \dd^4x \sqrt{- g}
\[ \frac12 \mpl^2  R + \L_{\rm SM}+\L_{\rm KI}\].
\label{Ltot}
\ee
with
\be\label{master1}
\mathcal{L_{\rm{KI}}}=\frac{ G^{\mu\nu}}{M^2} D_\mu \mathcal{H}^\dagger D_\nu \mathcal{H} + \sum_a \frac14  \alpha_A 
\frac{3 \G^{\mu\nu \alpha\beta}}{M^2}
F^a_{\alpha\beta}F^a_{\mu\nu}+\sum_i \alpha_\psi \frac{ G^{\alpha\beta}}{M^2}   \psi_i
i\gamma_\alpha D_\beta \psi_i,
\ee
with $\mathcal{H}$ is the Higgs doublet, and $ R, \, G^{\mu\nu}, \, {\mathcal{G}}^{\mu\nu \alpha\beta}$ the Ricci scalar, the Einstein tensor, and the double-dual Riemann tensor respectively.  The summation in the gauge and fermion terms runs over the SM gauge groups and fermions respectively.  For simplicity we take the non-minimal couplings for the gauge (fermion) fields universal, i.e. the same for all gauge groups (fermions), the results can easily be generalized.  With just a constant value of the non-minimal Higgs coupling this is the original version of the new Higgs inflation proposal of Ref. \cite{Germani:2010gm}, while non-minimal couplings for other sectors were introduced in \cite{Germani:2011cv,disformal}.  We parameterize the non-minimal gauge boson and fermion couplings as
	\be
	\alpha_i  = \alpha_{0i} \delta^{n_i/2}
	\label{alpha}
	\ee
	for $i=A,F$ and $\delta$ defined as
	\be\label{delta}
	\delta \equiv \frac{V}{\M^4} \simeq\frac{\lambda( \mathcal{H}^{\dagger}\mathcal{H})^2}{{\M^4}},
	\ee
        where we introduced the scale $\M^2 = M \mpl$, and with $V$ the Standard Model Higgs potential, where we dropped the quadratic term that plays no role at large field values. The couplings vanish for $\alpha_{0i} =0$ and are constant for $n_i=0$, which are arguably the most interesting cases.
Thus, in the following we study four cases:
\begin{align}
{\rm Case} \; {\rm A}:& \;\; (\alpha_A,\alpha_f)=(0,0), \nn \\
{\rm Case} \; {\rm B}:& \;\;  (\alpha_A,\alpha_f) = (\alpha_{0A}=\mathrm{const},\,0), \nn \\
{\rm Case} \; {\rm C}:& \;\;  (\alpha_A,\alpha_f) = (\alpha_{0A}=\mathrm{const},\,\alpha_{0f}=\mathrm{const}), \nn \\
{\rm Case} \; {\rm D}:& \;\;  
(\alpha_A,\alpha_f)=(\alpha_{0A}\delta^{n_A/2},\,\alpha_{0f}\delta^{n_f/2}),\,\, \mathrm{with}\quad\,n_A\geq 1\,\, \&\, \,n_f\geq 0.
 \label{cases}  
\end{align}

\subsection{Standard Model with non-trivial kinetic sector}\label{model}
The Higgs-gravity sector can be brought in (approximate) standard form via a disformal transformation of the metric \cite{disformal}\footnote{Note that, contrary to the conformal transformation used in Higgs inflation, the disformal transformation leads to the same Lagrangian \cref{Lmodel} both in the metric and Palatini formulation of general relativity (at first order in $\varepsilon$).}:
\be
g_{\alpha\beta} \longrightarrow  g_{\alpha\beta}-\varepsilon_{\alpha\beta}
\equiv  g_{\alpha\beta}-2\frac{D_\alpha \Phi^\dagger	D_\beta \Phi}{\M^2}.
\label{disformal}
\ee
The transformation leads to the action 
\begin{align}
\label{Lmodel}
\mathcal{L} &=
-K_{\phi}(\mathcal{H}) |D_\mu \mathcal{H}|^2- V(\mathcal{H})-\sum_a \frac14 K_A(\mathcal{H})(F^a_{\mu\nu})^2 
              +\sum_i K_{\psi}(\mathcal{H})\bar \psi_i (i\slashed{D}) \psi_i
              \nn \\ &
- \Big(
  \frac{y_t}{\sqrt{2}} \bar q_L \mathcal{H}_c t_R L + {\rm h.c.}
  \Big).
\end{align}
$F^a_{\mu\nu}$ runs over the SM gauge groups, and $\psi_i$ over the left- and right-handed fermions.  We only added the Yukawa interaction for the top quark, as this gives the dominant contribution to the running, with $q_L$ the left-handed doublet, $t_R$ the right-handed top quark and ${ \mathcal{H}}_c =(i\sigma^2)\mathcal{H}^{*}$. For simplicity from now on $y_t\equiv y$. The non-minimal Higgs, gauge and fermion field space metrics are given explicitly by
\begin{align}\label{functions}
K_{\phi} &= \(1+ \delta \), &
K_{A} &= \(1+ \alpha_A \delta\) ,&
K_{\psi} &= \(1+ \alpha_F \delta\) .
\end{align}
Eqs. \eref{Ltot} and \eref{Lmodel} are equivalent up to $\O(\varepsilon, \eps)$ corrections, with $\varepsilon$ defined in \eref{disformal} and $\eps = - \dot H/H^2 $.  During inflation $\varepsilon \sim \eps \ll 1$ are slow-roll suppressed, however after inflation the corrections can become large \cite{Fumagalli:2017cdo}. These corrections are degenerate with the quantum corrections in the mid-field regime, and will be aborbed in the threshold corrections that we will introduce in \cref{pred}.  The SM Higgs doublet can be parameterized
\be
{\cal H} = \frac1{\sqrt{2}} \left(
  \begin{array}{c}
    \theta_1+ i \theta_2 \\ \vp + i \theta_3
    \end{array}
  \right)
= \frac1{\sqrt{2}} \left(
  \begin{array}{c}
    \theta_1+ i \theta_2 \\ \phi + \delta \phi + i \theta_3
    \end{array}
  \right)
  \label{H_doublet}
\ee
with $\phi(t)$ the classical background and $\delta\phi(x,t)$ and $\theta_i(x,t)$ the Higgs and Goldstone fluctuations.

From now on we take \cref{Lmodel} as our starting point. To avoid notational clutter, if there is no chance of confusion we will suppress the flavor indices on the fermion, gauge and Goldstone fields, and on the Yukawa and gauge couplings.

\subsection{Tree level analysis and different regimes}\label{largefieldsection}

The action for the classical background Higgs field $\phi$ in \cref{H_doublet} becomes
\be
\mathcal{L}=\frac{1}{2}\mpl^2 R
- \frac12K_{\phi} (\phi) (\partial_\mu \phi)^2
-\frac{\lambda}{4} \phi^4 .
\label{L_inf}
\ee
The dynamics of the system is very different for small and large field values, for which the correction to the Higgs kinetic term is not important respectively dominates. We use the parameter $\delta$ in \cref{delta}  evaluated on the background to parameterize the diffferent regimes:
\begin{equation}
\rm{small\,field\,:\quad}\delta\ll 1,\quad\quad\rm{large\,field\,:\quad}\delta\gg 1.
\end{equation} 
In the small field limit the action reduces to the SM Lagrangian while the large field regime corresponds to the inflationary regime.

Consider the latter and take $\delta \gg 1$.  In terms of canonically normalized background field $h$, defined via $\partial h =\sqrt{K_\phi}\partial \phi$, the potential becomes
\be
V=\frac{\tilde{\lambda}}{4} \mpl^4  \(\frac{h}{\mpl}\)^{4/3},\quad \tilde{\lambda} =6^{4/3} \lambda^{1/3} 
\(\M/\mpl\)^{8/3}.
\label{chaL}
\ee
which is chaotic inflation with exponent of $4/3$ in the potential. It is easy to check that the full period of inflation happens well inside the large field regime.  Taking $N_* =60$ for the CMB pivot scale, the spectral index and tensor-to-scalar ratio at tree level are
\be  
n_{s}=1+\frac{d\ln \mathcal {P}_{R}}{d\ln k}\Big{|}_{\star}  \simeq 0.972,\qquad r=\frac{\mathcal{ {P}}_{T}}{\mathcal {P}_{R}}\Big{|}_{\star} \simeq 0.089
\label{nsrtree}
\ee
where $\mathcal {P}_{R},\,\mathcal {P}_{T}$ are the scalar and tensor power spectrum and we use the standard single-field slow-roll approximation that leads to $n_{s\star}-1\simeq 2\eta_{V\star} -6\eps_{V\star}$ and $r\simeq16\eps_{V\star}$ with potential slow-roll parameters defined as $\eps_V=\mpl^2/2 (V'/V)^2$ and $\eta_{V}=\mpl (V'' /V)$. The scalar power spectrum fixes the free parameter $M \simeq 1.5 \times 10^{-8} \mpl \lambda^{-1/4}$. The aim of this paper is to calculate the sensitivity of these results to one-loop RG flow corrections.

\section{Covariant formalism}\label{covariantformalism}
We now calculate the RGEs in the inflationary regime.  We sketch the methodology and introduce the relevant notation, before presenting the results. The technical details are relegated to \cref{A:details}.

The covariant construction is based on the definition of a metric on the field space manifold that is defined by the tensor contained in the highest-derivative term of the action \cite{Vilkovisky:1984st}. 
This approach has led Vilkovinsky and de Witt in the Eighties to build the covariant effective action \cite{DeWitt:1965jb,Vilkovisky:1984st}, and it is ubiquitous to formulate multi-field dynamics in a covariant form \cite{Sasaki:1995aw,GrootNibbelink:2001qt,Achucarro:2010da,Gong:2011uw}.
Recently there has been a renewed interest in using geometric covariant formalisms for instance in the context of SMEFT (and extensions) \cite{Alonso:2015fsp,Alonso:2016oah,Helset:2018fgq,Nagai:2019tgi,Helset:2020yio}, to compute UV divergences of general relativity as an EFT \cite{Alonso:2019mok} and to address the issue of frame equivalence (at quantum level) in scalar-tensor theories \cite{Falls:2018olk,Finn:2019aip}.

Our work aims to provide covariant counterterms to have results that are field reparameterization independent at each step in the derivation of the beta functions. This allows to more systematically include all interactions contributing at a given order, and simplifies calculations considerably.

In truth, once gauge bosons are included, the desired full covariant result is obtained either by using a metric projected on the space of gauge orbits or by using the metric \cref{manifold} but with a specific gauge choice \cite{Fradkin:1983nw,DeWitt:1967ub} different from the Landau gauge used in this work.  We consider this delicate issue in detail (and for a more general set-up) in a forthcoming publication that will make it clear that results obtained by means of the two different gauge choices (and the same field space metric in \cref{manifold}) differ by terms that are only higher order in the accuracy used in the large and small field regimes.
\subsection{Large field regime and counterterms}\label{largecounter}
In the large field regime we can use \eref{Lmodel} and expand in small $\delta^{-1}$ to capture the dominant effects.
At leading order in the $\delta^{-1}$-expansion the mass $\M$ is not an independent parameter, as it can be rescaled from the Lagrangian. Indeed, if we define the tilde fields and couplings via
\be
\tilde{\mathcal{H}}^6 = \frac{\lambda}{\M^4} \mathcal{H}^6, \qquad
\tilde{\lambda} =6^{4/3} \lambda^{1/3} 
\M^{8/3}, \qquad
\tilde y = y \(\frac{\M^4}{\lambda}\)^{1/6},\quad
\tilde \alpha_{0i} = \alpha_{0i} \(\frac{\lambda}{\M^4} \)^{\frac13(1+\frac{n_i}{2})}
\label{rescale}
\ee
the Lagrangian in the large field regime becomes
\begin{align}
\label{L_rescale}
\mathcal{L} &=
-\tilde {\mathcal{H}}^4 |D_\mu \tilde {\mathcal{H}}|^2- 6^{-4/3} \tilde \lambda |\tilde {\mathcal{H}}|^4
- \Big(
  \frac{\tilde y}{\sqrt{2}} \bar q_L \mathcal{H}_c t_R + {\rm h.c.}
  \Big)
\nn \\
& -\sum_a \frac14 \tilde \alpha_{0A} |\tilde {\mathcal{H}}|^{4(1+\frac{n_A}{2})}  (F^a_{\mu\nu})^2 
+ \sum_i \tilde \alpha_{0F} |\tilde {\mathcal{H}}|^{4(1+\frac{n_F}{2})}\bar \psi_i (i\slashed{D}) \psi_i + \mathcal{O}(\delta^{-1}).
\end{align}
Also in the small field regime, where the model reduces to the SM, the scale $\mathcal{M}$ drops out of the Lagrangian at leading order in the $\delta$-expansion. The scale $\M$ cannot be removed over the whole field range though.  In fact, it still plays a fundamental role in determining the boundaries between the small and large field region and the matching conditions between the parameter of the two (see \cref{matching}).

We will calculate the loop corrections in the untilde variables, as this makes the $\delta^{-1}$ -expansion more transparent. As just shown, the original variables are not all independent at first order in the $\delta^{-1}$           expansion, and the resulting counterterms will form a system that is not closed.\footnote{If one tries to find the running of $\mathcal{M}$ (and the other original parameters in the Lagrangian), there are not enough conditions to solve for all counterterms and derive all beta functions.}  We then translate the counterterms to those for the set of independent tilde-variables \eref{Ztilde} to derive the beta functions for the tilde variables.

Counterterms are introduced in the usual way by rescaling the bare fields and couplings by $Z_i =1 + \delta_i$, with $\delta_i$ the counterterm:
\begin{align}
\mathcal{H}_b &= \sqrt{ Z_\phi}  \mathcal{H}, &
\psi_b &= \sqrt{Z_{\psi} } \psi,&
A^\mu_b& = \sqrt{Z_A}  A^\mu,&
\alpha_b &= Z_{\alpha} \alpha,
\nn \\
\lambda_b &= Z_\lambda \lambda, &
\M_b &= Z_\M \M, &
y_b &= Z_y y, &
g_b &= Z_g g.
\label{Zelement}
\end{align}
The relation to the tilde counterterms follows from the rescaling relation \cref{rescale} and is
\be
Z_{\tilde \phi}^3 = Z_\lambda Z_\M^{-4} Z_\phi^3,\quad
Z_{\tilde \lambda} = Z_\lambda^{1/3} Z_\mathcal{M}^{8/3} , \quad
Z_{\tilde y} = Z_y Z_\M^{2/3} Z_\lambda^{-1/6},\quad
Z_{\tilde \alpha_i} =Z_{\alpha_i} (Z_\lambda Z_\M^{-4})^{\frac13(1+\frac{n_i}{2})}.
\label{Ztilde}
\ee
To extract the beta functions only the divergent part of the one-loop corrections is needed. We will calculate the two-point functions using dimensional regularization where we drop all finite contributions and only keep the $\epsilon$-poles.  The one-loop beta functions for the couplings and anomalous dimensions are then extracted from the counterterms via
\be
-\partial_t (\ln Z_{\lambda_{i}}) =\lim_{\epsilon \rightarrow 0}\eps (Z_{\lambda_{i}}-1) = \frac{\beta_{\lambda_{i}}}{\lambda_{i}}, \qquad
\frac12 \partial_t (\ln Z_\phi)=-\frac12\lim_{\epsilon\rightarrow 0}\eps(Z_\phi-1) =\gamma_\phi,
\label{def_beta}
\ee
with $t = \ln \mu$, $\mu$ the renormalization scale 
and $\epsilon=4-d$.

\subsection{Covariant fields}\label{covariant fields}

The Higgs field is decomposed in background plus perturbations as in \cref{H_doublet}.
We work in Landau gauge (see \cref{L_GF}) for which the ghosts fields decouple.   The bosonic fields can be grouped together $\varphi^{I}=\phi^{I}+\delta\phi^{I}=\{\varphi^a,A_\mu^i\}$, with $a$ running over the Higgs and Goldstone fields $\vp^a=\{\varphi, \theta_1,\theta_2, \theta_3\}$, and $i$ running over the SM gauge fields. The field space manifold has a non-trivial geometry defined by the metric
\begin{equation}\label{manifold}
G_{IJ}=\{K_{\phi}(\varphi^{I})\delta_{ab},K_{A_j}(\varphi^I)\eta_{\mu\nu} \delta_{ij}\}=\{(1+\delta)\delta_{ab},(1+\alpha_A\delta)\eta_{\mu\nu}\delta_{ij}\}.
\end{equation} 
Because the fluctuations $\delta\phi^I$ are not covariant objects on the field space manifold, one has to deal with intermediate results (for the counterterms and the effective action) that are not covariant under fields reparameterizations. This makes it hard to organize the calculation and include all relevant interactions. For example, it is well known that the one-loop Coleman-Weinberg potential depends on the covariant mass matrix of the bosonic fields running in the loop $m^I_J= G^{IK}\del_K \del_J V$, with $\del_I$ the covariant derivative constructed from the field space metric in \cref{manifold}.  To obtain this result expanding the action in $\delta \phi^{I}$ requires using the background equations of motion as well.

To expand the Lagrangian in a form that is fully covariant under field redefinitions one should replace the ordinary field displacement $\varphi^I-\phi^I$ with the tangent vector to the unique geodesic connecting the background $\phi^I$ to the field $\varphi^I$ 
\begin{equation}
\varphi^I-\phi^I\rightarrow Q^I=\frac{d\varphi^I(\tau)}{d\tau}\big|_{\tau=0},
\end{equation}
where $\tau$ is the affine parameter parameterizing the geodesic such that $\varphi^I(\tau=1)=\varphi^I$ and $\varphi(0)=\phi^I$. Let us summarize the notations in the following table
\begin{center}
	\begin{tabular}{| l | r l |}
		
		I & Non-covariant fields $\delta\phi^{I}$ \Big| & Covariant fields $Q^{I}$  \\
		\hline
		$\phi$ & $\delta\phi\quad =$ & $Q^{\phi}+\mathcal{O}(Q^2)\quad$\\
		
		$\theta$ & $\theta \quad=$ & $Q^{\theta}+\mathcal{O}(Q^2)$ \\
		
		A & $A_{\mu} \quad=$ & $Q^{{A}_{\mu}}+\mathcal{O}(Q^2)$  \\
\label{T:notation}		
	\end{tabular}
\end{center}
In order to find the relation between the non-covariant displacements $\delta\phi^I$ and the covariant ones $Q^I$, one can expand $\delta\phi^i\equiv\varphi^i-\phi^i\equiv\varphi^i(\tau=1)-\varphi^i(\tau=0)$ in Taylor series around zero in the affine parameter and recursively  use the geodesic equation satisfied by $d\varphi^I/d\tau$; this gives 
 \begin{equation} \delta\phi^i=\sum_{n=0}^{\infty}\frac{1}{n!}\frac{d^n\varphi^i}{d\tau^n}\Big|_0=Q^i-\frac{1}{2}\Gamma^i_{jk}Q^jQ^k+\frac{1}{3!}(\Gamma^i_{lm}\Gamma^m_{jk}-\partial_l \Gamma^i_{jk})Q^jQ^kQ^l+....
 \label{covnoncov}
 \end{equation}
 where $\Gamma_{IJ}^K$ are the Christoffel symbols associated to the metric $G_{IJ}$ evaluated on the background.
 To expand the action in covariant form we can consider $S(\varphi)$ as a function of the affine parameter $\tau$ evaluated in $\tau=1$, i.e.
\begin{equation} 
S(\varphi)=\sum_{n=0}^{\infty}\frac{1}{n!}\frac{d^{n}S}{d\tau^{n}}\Big|_{\tau=0}=\sum_{n=0}^{\infty}\frac{1}{n!}Q^{i_{1}}\cdot\,\cdot\,\cdot Q^{i_{n}}[\nabla_{(i_{1}}\cdot\,\cdot\,\cdot\nabla_{i_{n})}S][\phi],
\label{taylor}
\end{equation} 
where we used $ \frac{d}{d\tau}\equiv\frac{d\varphi^i}{d\tau}\nabla_{i} $ and the geodesic equation.  The round brackets mean symmetrization over the indices.  The coefficients of the expansion are evaluated on the background and will determine the strength of the interactions. In particular, we expand in this way the scalar functions $\{V,K_{\phi},K_{\psi},K_{A}\}$ and the Yukawa and gauge interactions.  Equivalently, the action can be expanded by normal Taylor series in the fluctuations $\delta\phi^I$, and then substitute their expression in terms of the covariant fields given in \cref{covnoncov}.\footnote{The two procedures give the same results since both represent the same expansion of the action in the affine parameter $\tau$.} We use this second approach to expand the kinetic terms in covariant fluctuations.  In this expansion we neglect terms with time derivative of the background ($\dot \phi^2$-corrections), as well as the backreaction from gravity, which are both slow-roll suppressed during inflation.

\section{Renormalization group flow}
\label{s:renormalizability}

In this section we calculate the one-loop beta functions for SM with non-minimally derivative couplings \cref{Lmodel}. In the small field regime the set-up reduces to the SM EFT with the SM RGEs to leading order in the $\delta$-expansion. In the large field inflationary regime, the EFT can be expanded in $\delta^{-1}$.  As we show, the EFT is renormalizable in the sense that all divergences can be absorbed in counterterms order by order in the $\delta^{-1}$-expansion.

\subsection{Renormalization group equations}

We compute one-loop corrections to the Higgs, Goldstone boson, fermion and gauge boson two-point functions, and expand in $\delta^{-1}$ to find the leading order contribution in the large field regime. This gives the various counterterms in the theory, and consequently the beta functions using \cref{def_beta}. The idea is to provide a systematic procedure to compute one-loop beta functions in similar set-ups. 
Let us remind that we calculate the quantum corrections in the untilde variables, as this makes the $\delta^{-1}$-expansion more transparent.  However, we rewrite the results in terms of the independent (and relevant) set of couplings, for which we derive the RGEs.

The momentum dependent part of the two-point functions gives the counterterm for the kinetic terms, whereas the momentum independent part provides the counterterm for the two-point vertexes.  We will denote these with $Z_{2f}$ and $Z_{c \,2f}$ respectively, with $f=\{Q^{\phi},Q^{\theta},Q^{A},\psi\}$ the (covariant) fields in question, and $c=\{\lambda,g,y\}$ if it renormalizes the Higgs, gauge or Yukawa coupling. The relevant counterterms are those of the quadratic Lagrangian. These $Z$-factors can be expressed in terms of the basis set of counterterms introduced in \eref{Zelement}.  For example, from the Higgs kinetic term in the large field regime
\be
\mathcal{L} 
\supset\frac12 Z_\phi^2 Z_\lambda Z_\M^{-4}Z_{\phi}\delta\,(\partial Q^{\phi})^2\equiv \frac12 Z_{2Q^{\phi}}\delta\,(\partial Q^{\phi})^2.
\ee
The full set of $Z$-relations in the large field regime and at leading order is given in \cref{Z2pnt}.

To understand the results given in the next subsections it is useful to look at the masses of the various particles. This allows to determine which particles have masses of the inflationary scale and are included in the EFT spectrum for different choices of the non-minimal couplings, and which are too heavy or too light (too weakly coupled) to contribute to the loop corrections at leading order.  The masses of the gauge and fermion fields depend on the functions $K_{A}$ and $K_{\psi}$ in their kinetic terms. The masses of the bosonic fields are given by the covariant expression $(m^2)_I^J = -G^{IJ} \nabla_I \nabla_J \mathcal{L}$ evaluated on the background:
\begin{align} \label{mass1}
m_h^2 & = \lambda \phi^2 \frac{(3+\delta)}{(1+ \delta)^2} , &
m_{\theta}^2 & = \lambda \phi^2 \frac{(1+3\delta)}{(1+ \delta)^2 }, &
m_{A}^2 & = \frac{g_i^2 \phi^2(1+\delta)}{(1+\alpha_A \delta)} +\frac{\delta(2+n_A) \alpha_A \lambda \phi^2}{(1+\delta)(1+\alpha_A \delta)},
\end{align}
with $n_A$ determining the non-minimal gauge coupling  \eqref{cases}.
We used the notation $(m^2)^{Q^{a}}_{Q^{a}}\equiv m^2_a$.  The last term in the gauge boson mass arises from mixing between the Higgs and gauge sector (specifically, because $\Gamma^\phi_{AA}\neq 0$), and it is suppressed at large field values for $n_A < 1$. In principle we should also define covariant fermion fluctuations, but it is not clear how to do that rigorously .  We can find a parametric estimate of the fermion mass by rescaling $K_{\psi}(\phi) \psi \to \psi$ to obtain approximately canonically renormalized fermions, where we evaluate the function $K_\psi$ on the background. This gives
\be\label{mass2}
m_{\psi} \sim
\frac{1}{\sqrt{2}} K_{\psi}^{-1} y_i \phi.
\ee 
Below we will summarize the results for the two-point functions and beta functions for the various cases defined in \cref{cases}. The technical details and the Feynman rules are given in \cref{A:details}.

\subsubsection{Case A}
\label{s:caseA}

Let us start with case A.  The top quark and gauge bosons are minimimally coupled and have standard kinetic terms for $\alpha_{0f} =\alpha_{0A}=0$.  The Higgs and Goldstone are light $m_{h,\theta}^2 \sim {\cal O}(\delta^{-1}) V^{1/2}$, and their fluctuations decouple; this holds for all the cases we discuss. The top quark has mass $m^2_t \sim V^{1/2}$ and is in the spectrum.  The gauge bosons on the other hand are heavy $m_A^2 =\O(\delta) V^{1/2}$ and should be integrated out; to obtain a renormalizable EFT at lower energies requires new physics at this mass scale, as the gauge field loop contribution is non-renormalizable \cite{Fumagalli:2017cdo}.\footnote{The loop-correction to the Higgs two-point vertex can be calculated in the $g \ll 1$ limit, such that the gauge boson mass is below the cutoff scale of the theory. It scales with $\delta^2$ and is thus large, and it cannot be absorbed in the counterterms of the Lagrangian. }

The one-loop expressions for the self-energies are given in appendix \ref{A:abel}-\ref{A:SM}.  The counterterms for the Higgs and fermion kinetic terms and for the fermion two-point interaction vanish, while the Higgs two-point vertex gets a corrections from the top loop:
\begin{align}
Z_{2Q^{\phi}} =Z_{2\psi} = Z_{y2\psi}  =1+ \mathcal{O}(\delta^{-1}) , \qquad
Z_{\lambda 2Q^{\phi}} =  1 + \boldsymbol{A} + \mathcal{O}(\delta^{-1})
\label{Z_caseA}
\end{align}
with
\be
\boldsymbol{A} = -\frac1{8\pi^2\eps}\( \frac{N_c y^4}{\lambda}\)=-\frac1{8\pi^2\eps}\( \frac{ 6^{4/3} \tilde y^4}{\tilde \lambda}\).
\ee
with $N_c =3$ the number of colors.

We can understand these results parameterically by setting $K_\phi$ to its constant background value, and evaluting diagrams with the unrenormalized Higgs field $\delta \phi$ and fermion fields on the external lines.  The effect of $K_\phi \sim \delta$ is that the Higgs/Goldstone propagator is suppressed by a factor $\delta^{-1}$, the Higgs-gauge couplings (which reside inside the kinetic terms) enhanced by a factor $\delta$, and the diagram with a $Z_{2Q^\phi}$ counterterm is also enhanced by a factor $\delta$. Thus Higgs loops are suppressed, and since the gauge fields are integrated out, there is no leading order contribution to the fermion self-energy. The top loop contribution to the Higgs two-point vertex is as in the SM, however the wave function correction  $Z_{2Q^\phi} \propto \delta^{-1}$ is suppressed. 

The beta functions are derived using \eref{def_beta}, and depend on the logarithm of the $Z$-factors. We are thus interested in
\begin{align}
0& =\ln (Z_{2Q^{\phi}}) 
= \ln(Z_{\tilde \phi}^3),&
0& =\ln(Z_{2\psi}) ,\nn \\
0&= \ln(Z_{y{2\psi}}) 
= \ln\(Z_\psi Z_{\tilde y} Z_{\tilde \phi}^{1/2} \),& 
\ln(1+ \boldsymbol{A}) &=\ln(Z_{\lambda 2Q^{\phi}}) 
= \ln\(Z_{\tilde \lambda} Z_{\tilde \phi}^2\),
\label{ZcaseA}
\end{align}
where in the second step we used the relation between different counterterms (see \cref{Z2pnt} in the appendix) and \eref{Ztilde}.  Note that $Z_\M$ has dissapeared when written in terms of the tilde-variables, as it should. We can solve this system of equations to get
\be
Z_{\tilde \phi} =Z_{\psi} = Z_{\tilde y} =1, \quad Z_{\tilde \lambda} = 1 +\boldsymbol{A}.
\ee
The beta functions are then
\be
(\beta_{\tilde y},\gamma_\psi,\gamma_{\tilde \phi}) =\mathcal{O}(\delta^{-1}), \qquad 
\beta_{\tilde \lambda} = -\frac1{8\pi^2} 6^{4/3} N_c \tilde y^4.
\ee

\subsubsection{Case B}
\label{s:caseB}

Consider non-minimal kinetic terms for the gauge fields $K_A =1 + \alpha_0 \delta$, which brings them back in the spectrum during inflation $m_A^2 \sim V^{1/2}$. The counterterms now become
\begin{align}
&Z_{2Q^{\phi}} =Z_\psi=Z_{m_\psi} = Z_{2Q^A} =Z_{g2Q^A} = 1+\mathcal{O}(\delta^{-1}),\\[1mm]
&Z_{\lambda 2Q^{\phi}}
= Z_{\lambda 2 Q^{\theta}}=Z_{V}=1 +\boldsymbol{A}
\label{Z_caseB}
\end{align}
with
\be
\boldsymbol{A} = \frac1{8\pi^2\eps} \frac1{\lambda} \(\sum_i \frac{3g_i^4}{\alpha_0^2} -N_cy^4 \)=
\frac1{8\pi^2\eps} \frac{6^{4/3} }{\tilde \lambda} \(\sum_i \frac{3 g_i^4}{\tilde\alpha_0^2} -N_c \tilde y^4 \).
\ee
The summation is over the massive electroweak gauge bosons, the $W^\pm$ bosons and the $Z$-boson, with couplings $g_i = \frac12 \times \{g,g,\sqrt{g^2 +g^{'2}}\}$, and $g$ and $g'$ the $SU(2)$ and $U(1)$ gauge coupling respectively.

The counterterms derived from the Higgs self-energy are consistent with those derived from the Goldstone self-energy, and also from the effective potential given in \cite{Fumagalli:2017cdo}. However, it is interesting to note that only by including some of the genuinely new interactions coming from the $Q$-expansion of the gauge fields (that are absent for canonical gauge fields), we are able to find agreement for the various counterterms. Particularly important are the interactions\footnote{The couplings $\mathcal{K}_{Q^{\phi} 2\partial Q^I} =\mathcal{K}_{Q^{\theta} 2\partial Q^I}  =0$ vanish.}
\be
\mathcal{L}_{\rm{k}} = -\frac12 G_{IJ}(\varphi^I)\partial\varphi^I\partial\varphi^J \supset - \mathcal{K}_{2Q^{\phi} 2 \partial Q^I} (Q^{\phi})^2  (\partial Q^I)^2 - \mathcal{K}_{2Q^{\theta} 2 \partial Q^I} (Q^{\theta})^2  (\partial Q^I)^2 
\ee
where $\mathcal{K}_{2Q^I \partial Q^J}$ are the background dependent coefficients given by expanding the non-canonical kinetic term in covariant fields. For example $\mathcal{K}_{2Q^{\phi} 2 \partial Q^A}=\alpha_{A0}\delta\{0,-1/3\phi^2\}$ where the terms in curly brackets give the leading terms in the SM regime and large field regime respectively.  These new interactions give a contribution to the Higgs self-energy
\begin{align}
\delta \Pi_{Q^{\phi}} =& \feynmandiagram [baseline=(a.base),horizontal=i1 to i3, layered layout] 
{ {i1} --  a [square dot, red] --  {i3}, 
	a -- [photon, out=135, in=45, min distance=1.8cm, edge label= $\small{Q^{A}}$] a ,
};+\feynmandiagram[baseline=(a.base),horizontal=i1 to i3, layered layout] 
{ {i1} --  a [square dot, red] --  {i3}, 
	a -- [dashed, out=135, in=45, min distance=1.8cm, edge label= ${Q^{\theta}}$] a,
};+\feynmandiagram [baseline=(a.base),horizontal=i1 to i3, layered layout] 
{ {i1} --  a [square dot, red] --  {i3}, 
	a -- [out=135, in=45, min distance=1.8cm, edge label= ${Q^{\phi}}$] a,
}; \nn\\[1.5mm]
  =& -\frac{1}{8\pi^2\eps}\sum_I 2n_I \textcolor{red}{ \mathcal{K}_{2Q^{\phi} 2 \partial Q^I} }G^{II}  m_I^4
     \label{Pi_new}
\end{align}
and similar for the Goldstone boson self-energy.  Here the sum is over $Q^I=\{Q^{\phi},Q^{\theta},Q^{{A}_\mu}\}$ with $n_I=\{1,3,4\times 3\}$ the d.o.f.  Only the gauge boson contributes at leading order, as the Higgs/Goldstone mass is suppressed during inflation. Including this correction, see appendix \eref{feynmannappendix} for more details, 
the results for all counterterms are consistent.

We can once again understand the results in \cref{Z_caseB} parametrically, by taking  $K_\phi,K_A \propto \delta$ on the background.  The difference with case A is that now also the gauge boson propagator is suppressed by $\delta^{-1}$, the gauge non-abelian self-interactions (which reside inside the gauge kinetic terms) enhanced by a factor $\delta$, and the diagram with a $Z_{2Q^A}$ counterterm is also enhanced by a factor $\delta$. Remembering that the gauge-Higgs couplings are enhanced, we see that now the gauge loop does contribute to the Higgs self-energy at leading order as the relevant diagrams scale with powers of $(K_\phi/K_A) = {\cal O}(1)$. There is no leading order contribution to $Z_{2Q^A}$. This is because both the diagrams with a gauge-loop (the enhancement of the gauge-interactions is cancelled by the suppressed gauge boson propagator) and fermion loop (the fermion interactions are standard) are $ {\cal O}(1)$, while the diagram with the counterterm is enhanced. There is also no gauge boson contribution to the fermion self-energy as the corresponding diagrams are suppressed by the gauge-boson propagator.

The Higgs and fermion $Z$-factors give the same results as in case A given in \eref{ZcaseA}, except that $\boldsymbol{A}$ now includes the gauge contribution. In addition, the gauge interactions give
\begin{align}
0 = \ln( Z_{2Q^A}) &
=\ln\(\frac{ Z_{\tilde \alpha_0} Z_{\tilde \phi}^2}{Z_g^2}\), \nn \\
0 = \ln(Zg_{2Q^A}) &
= \ln \(Z_{\tilde \phi}^3\)
\end{align}
where we used the Ward identity $Z_{g_i}^2 Z_{A_i} =1$ (no summation).  
We can solve the system of equations to get
\be
Z_{\tilde \phi} =Z_{\psi} = Z_{\tilde y} = \frac{Z_g^2}{Z_{\tilde \alpha_0}} =1+\mathcal{O}(\delta^{-1}), \quad Z_{\tilde \lambda} = 1 +\boldsymbol{A}.
\ee
The beta functions are then
\be
\beta_{\tilde y} =\beta_{g^2/\tilde \alpha_0} =\gamma_\psi = \gamma_{\tilde \phi} =\gamma_{A} =O(\delta^{-1}), \qquad 
\beta_{\tilde \lambda} = \frac1{8\pi^2} 6^{4/3}\( 3\sum_i \frac{g_i^4}{\tilde \alpha_0^2}- N_c\tilde y^4\), 
\ee
%

\subsubsection{Case C and  D}

Case C is analogous to case B with the only difference that this time the fermion is light and decouples. Thus $\beta_{\tilde \lambda} = 1/8\pi^2 6^{4/3}( \sum_i 3g_i^4/\tilde{\alpha}_0^2)$. In case D all fields are weakly coupled,  $\beta_{\tilde \lambda} = \mathcal{O}(\delta^{-1})$, and nothing runs. 

\subsection{Beta-functions summary}

The RGEs during inflation can be summarized as follows
\begin{align}
\beta_{\tilde y} &=\beta_{g^2/\tilde \alpha_0} =\gamma_\psi = \gamma_{\tilde \phi} =\gamma_{A} = \mathcal{O}(\delta^{-1}),
\nn \\                     
\beta_{\tilde \lambda} &= \frac1{8\pi^2} 6^{4/3}\( 3 {f_1}\sum_i\frac{g_i^4}{\tilde \alpha_0^2}- {f_2} 3\tilde y^4\),
\label{inflationRGE1}                           
\end{align}
with  $g_i = \frac12 \{g,g,\sqrt{g^2+{g'}^2}\}$, and $g$ and $g'$ the $SU(2)$ and $U(1)$ gauge coupling respectively.
The $f_i$ take on a value of zero/one depending on the non-minimal couplings of the fermion/gauge fields, specifically
\begin{align}
{\rm Case} \; {\rm A}: \;\; (f_1,f_2) &= (0,1), \nn \\
{\rm Case} \; {\rm B}: \;\;  (f_1,f_2) &= (1,1), \nn \\
{\rm Case} \; {\rm C}: \;\;  (f_1,f_2) &= (1,0), \nn \\
{\rm Case} \; {\rm D}: \;\;  (f_1,f_2) &= (0,0).
\label{inflationRGE2}  
\end{align}

\section{Predictions for inflation}\label{pred}

We are now in the position to calculate corrections to the inflationary observables $n_s$ and $r$ due to the running of the couplings. 

For the boundary values of the couplings at the electroweak scale we use the two-loop matching conditions of \cite{RGE}.\footnote{The instability of the Higgs potential (when $\lambda(\mu)$ becomes negative) is pushed to larger scales with two-loop matching conditions compared to one-loop matching.} The inflationary parameters are sensitive to the top and Higgs masses $(m_t^{EW},m_h^{EW})\equiv(M_t,M_h)$ and the strong coupling constant $\alpha_s$. The top mass determination gives the largest uncertainty, both from an experimental and a theoretical point of view (see \cite{Hoang:2014oea,Nason:2017cxd} for recent discussions). The best current estimate is $M_t=172.9\pm0.4\,\mathrm{GeV}$ \cite{Tanabashi:2018oca}. We fix, for illustrative purposes, $\alpha_s$ to its central value $\alpha_s=0.1181$ and $M_h=125.6 \,\mathrm{GeV}$ \cite{Tanabashi:2018oca}.

In the small field regime ($\delta<1$) we use the the SM two-loop beta functions \cite{Luo:2002ey,RGE} to compute the running of the SM parameters $\{\lambda,y,..\}$.
At the boundary $\delta= 1$ we match to the rescaled couplings $\{\tilde{\lambda},\tilde{y},..\}$ of the large field regime $(\delta > 1)$, and use the one-loop beta functions derived in the previous section for their running. 
As already mentioned, the corrections due to unknown UV physics on the running gives rise to threshold corrections which we parameterize, following the approach in \cite{shap,critical1,Enckell:2016xse,Fumagalli:2017cdo}, by a jump in the couplings at the boundary between the two regimes.

We restrict to the tree-level RG improved potential for our numerical analysis since we have derived only the one-loop beta functions in the large field regime.

\subsection{Renormalization group dependence: analytical estimate}\label{s:flat}

Before turning to the actual numerical implementation we quickly recap the  analytical estimate of the effect of the running on the inflationary observables \cite{Fumagalli:2017cdo}. With the explicit form of the RGEs we can now verify some of the assumptions previously made.

\begin{figure}[t]
	\begin{subfigure}{1\textwidth}
		\centering
		\includegraphics[width=0.47\textwidth]{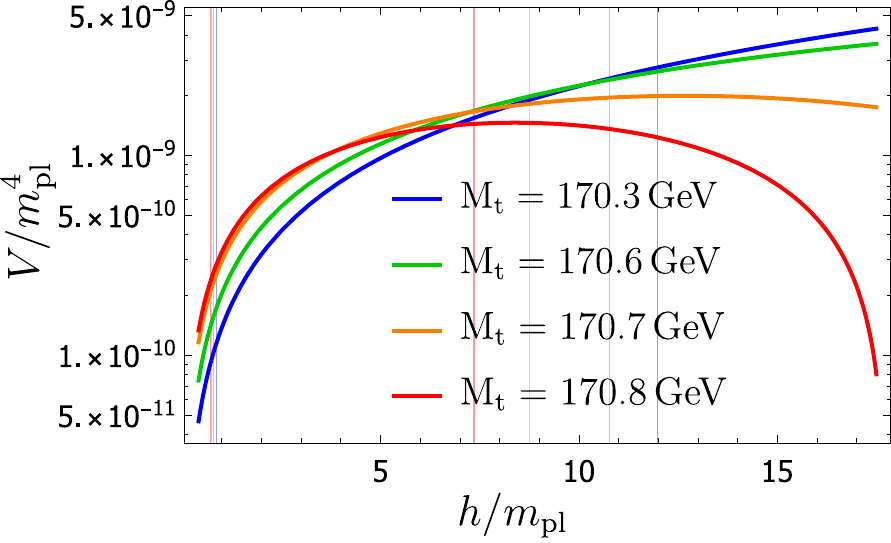} \hspace{0.5cm}
		\includegraphics[width=0.47\textwidth]{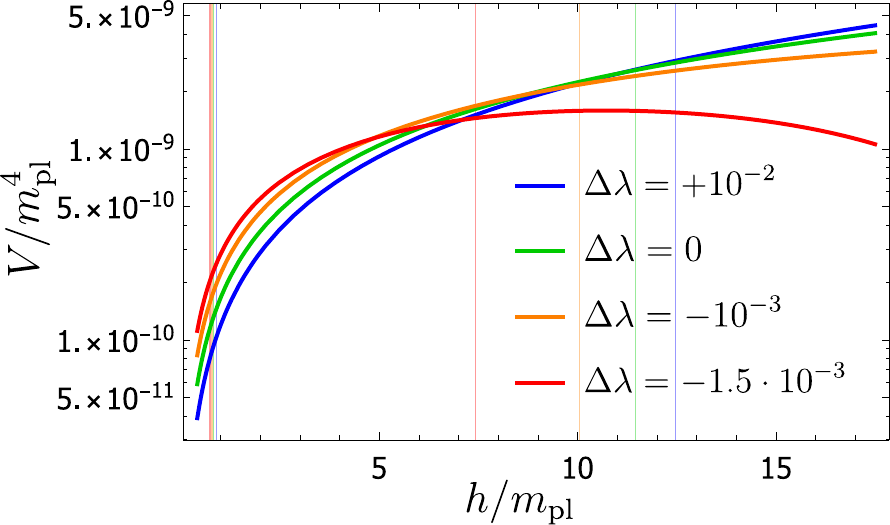} \hspace{0.5cm}
		\vspace*{-0.2cm}
		\caption{\textit{
				RG improved potentials as a function of the canonical inflaton field $h$.  The left (right) vertical lines correspond (for each scenario) to the beginning (end) of inflation.}}
		\label{fig1}
	\end{subfigure} \\
	\begin{subfigure}{1\textwidth}
		\centering
		\vspace*{0.3cm}
		\includegraphics[width=0.47\textwidth]{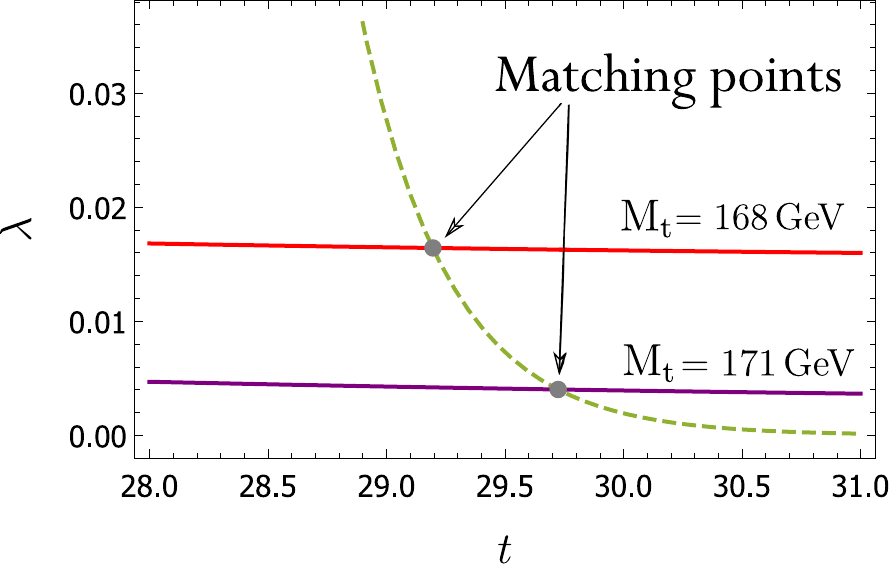}
		\includegraphics[width=0.47\textwidth]{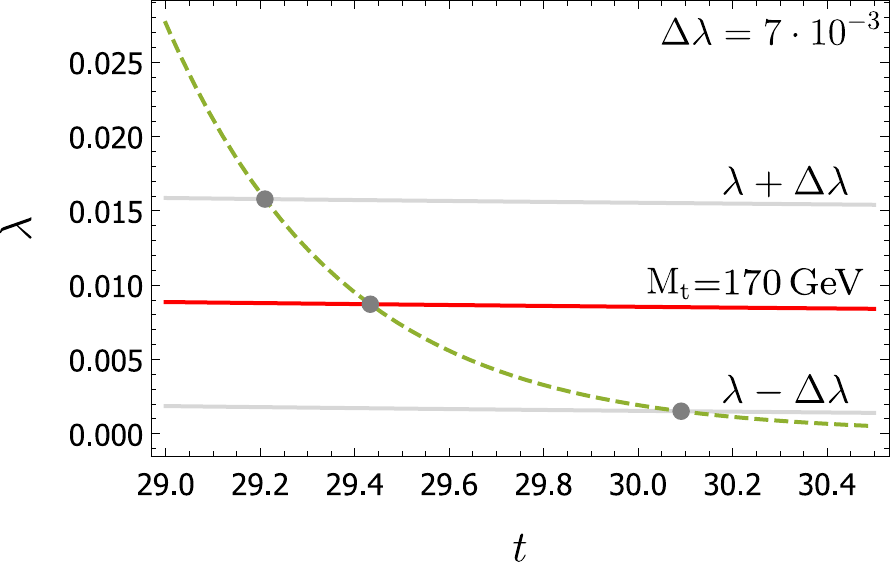}
		\vspace{.1cm}
		\caption{ 
			\textit{Intersection of the running coupling $\lambda(t)$ with the curve $ \alpha e^{-8t/3\,}$ (dashed green line) from  \eref{intersection} as a function of the renormalization time $t$. The intersection points give $(t_{\rm{eq}},\lambda_{\rm{eq}})$ and define the matching point. A larger top mass/negative threshold correction implies a smaller $\lambda_{\rm{eq}}$, and consequently larger corrections to the observables (see \cref{correction}).}}
		\label{texplanation}
	\end{subfigure}
	\caption{\textit{Case A: RG improved potential and matching point for $M_h=125.6\,\rm{GeV}$ and different top mass (left) and $(M_h,M_t)=(125.6,\, 170)\,\rm{GeV}$ and different threshold corrections (right).}}
\end{figure}

We consider the RG improved effective action in terms of the canonical field $h$, see \cref{chaL,rescale}
\be\label{cane2}
h=\frac{\sqrt{\lambda}\phi^3}{6\mathcal{M}^2}=\frac{\tilde{\phi}^3}{6},
\ee
frow which it follows that $\gamma_h=3\gamma_{\tilde{\phi}}$. In all cases A-D considered in the previous section we found $\gamma_{\tilde{\phi}}=\mathcal{O}(\delta^{-1})$, and we can thus neglect the anomalous dimension since $\gamma_h=\mathcal{O}(\delta^{-1}),\gamma'_h=\mathcal{O}(\delta^{-1})$.\footnote{As a consequence, our results for slow-roll parameters and observables are implicitly formulated in a gauge-invariant fashion as in \cite{Urbano:2019ohp}.} The leading-order RG improved action then becomes
\be
\mathcal{L}=\frac{1}{2} R -\frac{1}{2}\partial_\mu
h \partial^\mu h
-\frac{\tilde{\lambda}(t)}{4} h^{4/3}.
\label{chaL1}
\ee
The RG improved potential is shown in \cref{fig1} for different boundary conditions of the top mass and for different threshold conditions paramerized by a jump $\Delta \lambda$ in the Higgs coupling at the matching scale. For large top mass and/or large negative jump the potential develops a maximum.

We choose as renormalization scale the top mass. The RG time is then (using again \cref{rescale})
\be
t=\ln\left(\frac{\mu}{m_{{\rm t}}^{{\rm EW}}}\right),\qquad
\mu = y\phi =\tilde{y}\tilde{\phi}=\tilde{y\,}6^{1/3}h^{1/3}.
\label{choice}
\ee
Often the Yukawa coupling is neglected for simplicity as $y = \mathcal{O}(1)$. However, one should keep $\tilde y$ explicitly as it can be small. Indeed, using tree-level relations derived below \cref{nsrtree} for e.g. $y\simeq10^{-1}$ and $\lambda\simeq10^{-4}$ we find $\tilde{y}\simeq10^{-4}$.  Running effects enter the observables, because calculating slow-roll parameters (derivatives of the potential) also involves taking the derivative of the $\tilde \lambda$-coupling in the potential \cite{Fumagalli:2017cdo}:
\be
\frac{d \tilde{\lambda}(\mu)}{d h}
\equiv \beta_{\tilde \lambda} \frac{d t}{d h} = \frac{\beta_{\tilde \lambda}}{3h}+ \mathcal{O}(\delta^{-1}) ,\qquad
\frac{d\beta_{\tilde \lambda}}{dt}\equiv\beta_{\tilde \lambda}'=\mathcal{O}(\delta^{-1}),
\label{dlambda}
\ee
where we used the RGEs summarized in \cref{inflationRGE1}.
When the top quark decouples and the gauge boson remains in the spectrum, as in case C, the gauge boson mass is the appropiate scale and the same equations hold with $\tilde{y}_t$ replaced by $g/\tilde{\alpha}_0$.
The potential slow-roll parameters are
$
\eps_V= 8/(9 h^2)(1+\beta_{\tilde\lambda}/4\tilde\lambda)^2$ and
$\eta_V = 4 /(9 h^2)(1+5\beta_{\tilde{\lambda}}/4\tilde{\lambda})$. The number of $e$-folds
before the end of inflation is $N_\star \approx 3 h_\star^2D_\star/8$, where we {\it assumed}\footnote{Note that, in contrast to the case of the Cosmological Attractors (see \cite{Fumagalli:2016sof}), one cannot consider the beta functions dependent factor $D$ constant over the integration domain when computing $N_*$ as a function of the field. In fact, expanding the integrand, all the terms of the series contribute at the same order in the small parameter $\delta^{-1}$.}
\begin{equation}
D_\star=\(1+\frac{\beta_{\tilde\lambda}}{4\tilde\lambda}\)_\star^{-1} \approx {\rm constant}
\quad \Rightarrow \quad
\frac{\beta_{\tilde{\lambda}}}{{\tilde{\lambda}}}\ll \(\frac{\beta_{\tilde{\lambda}}}{{\tilde{\lambda}}}\)^2.
\end{equation}
To leading order in the $1/N_\star$ expansion the observables become
\be
n_s-1 \simeq -\frac5{3N_\star}\(1+\frac{2}{5}\frac{\beta_{\tilde\lambda}}{4\tilde\lambda}\)_\star,
\quad
r\simeq \frac{16}{3N_\star} \(1 +\frac{\beta_{\tilde\lambda}}{4\tilde\lambda}\)_\star
\label{observables}
\ee
The influence of the RG flow can become significant if the ratio $\beta_{\tilde \lambda}/(4{\tilde \lambda})$ is order one during inflation.

\subsection{Matching and running: numerical results for case A}\label{matching}

The linear analysis of the previous section indicates an RG dependence of the inflationary observables. We now compute numerically the size of these corrections, starting with case A.  In the small field regime we run the SM parameters $\{\lambda,y,..\}$ using the 2-loop SM beta functions \cite{Luo:2002ey,RGE}, and with boundary condtions at the EW scale. At the border between the two regimes
\begin{equation}
\delta|_{t_{{\rm eq}}}=1\implies\phi_{{\rm eq}}=\left(\frac{4\mathcal{M}^{4}}{\lambda(t_{{\rm eq}})}\right)^{1/4},\label{delta1}
\end{equation}
we switch to the rescaled couplings $\{\tilde{\lambda},\tilde{y},..\}$. Threshold corrections can be incorporated (and parameterized) by a jump in the coupling constants at $t_{\rm{eq}}$. Beyond the matching point we run with the one-loop beta functions valid in the large field regime \cref{inflationRGE1}.  The power spectrum constraint fixes
\begin{equation}
P_{\mathcal{R}}=2\cdot10^{-9}\implies \tilde \lambda_\star \equiv \tilde{\lambda}(t_{\star})=4\cdot10^{-10},\label{fixedpower}
\end{equation}
where $t_{\star}$ is as usual the value of the RG time \cref{choice} at horizon crossing.

The large field RGEs (\ref{inflationRGE1}\,,\ref{inflationRGE2}) for case A are
\begin{equation}
\beta_{\tilde{y}}\approx0,\quad\quad\frac{\beta_{\tilde{\lambda}}}{\tilde{\lambda}}=-\frac{3 \times 6^{4/3}}{8\pi^{2}}\frac{\tilde{y}^{4}}{\tilde{\lambda}}.\label{betatilde}
\end{equation}
The first equation trivially implies $\beta'_{\tilde{\lambda}}\approx0$. The matching conditions at the boundary depend on the scale $\mathcal{M}$
\begin{equation}\label{matchingcondition}
\tilde{\lambda}_{{\rm eq}}=6^{4/3}\lambda_{{\rm eq}}^{1/3}\mathcal{M}^{8/3},\qquad
\tilde{y}_{{\rm eq}}=y_{{\rm eq}}\lambda_{{\rm eq}}^{-1/6}\mathcal{M}^{2/3},
\end{equation}
where we used \eref{rescale}, and we introduced the notation $X(t_{{\rm eq}})\equiv X_{{\rm eq}}$.  We can now understand how the predictions for $n_{s}$ and $r$, which depend on the ratio $(\beta_{\tilde{\lambda}}/\tilde{\lambda})_{\star}$ with $\lambda_{\star}$ fixed, depend on the running. Different boundary conditions at the EW scale will result in different values of $\mathcal{M}$ required to adjust the matching conditions at $t_{{\rm eq}}$ in such a way that $\tilde{\lambda}_{\star}=4\cdot10^{-10}$ is obtained. Furthermore, different values of $\mathcal{M}$ (and $\lambda$) at the matching point will give a different value for $\tilde{y}_{\rm eq}$. Since $\beta_{\tilde{y}_{t}}\approx0$, $\tilde{y}_{\rm eq}=\tilde{y}_{{\rm \star}}$  this value will determine the correction to the inflationary parameters.

Given the simple form of \cref{betatilde} we can integrate $d\tilde{\lambda}/dt=\beta_{\tilde{\lambda}}$
explicitly 
\begin{equation}
\tilde{\lambda}(t)=\tilde{\lambda}_{{\rm eq}}+(t-t_{{\rm eq}})\beta_{\tilde{\lambda}}=\tilde{\lambda}_{{\rm eq}}+\ln\left(\frac{\tilde{y}\tilde{\phi}}{y_{{\rm eq}}\phi_{{\rm eq}}}\right)\beta_{\tilde{\lambda}}.
\end{equation}
It is possible to express $\tilde{\lambda}$ as a function of the field $h$ and the low energy parameters at the matching point $t_{{\rm eq}}$. Using \cref{betatilde,matchingcondition} we have
\begin{equation}
\tilde{\lambda}(h,t_{{\rm eq}})=6^{4/3}\lambda_{{\rm eq}}^{1/3}\mathcal{M}^{8/3}+\beta_{\tilde{\lambda}} (t_{{\rm eq}})\ln\left(\frac{h^{1/3}y_{{\rm eq}}\lambda_{{\rm eq}}^{-1/6}\mathcal{M}^{2/3}6^{1/3}}{m_{t}^{{\rm EW}}e^{t_{{\rm eq}}}}\right),\label{lat}
\end{equation}
where we used $y_{{\rm eq}}\phi_{{\rm eq}}=m_{t}^{{\rm EW}}e^{t_{{\rm eq}}}$ and \eref{choice}. Once we determine $t_{{\rm eq}}$, the corrections to $n_{s}$ and $r$ proportional to $\beta_{\tilde{\lambda}} (t_{{\rm eq}})/\lambda_{\star}$ can be computed. The value of \cref{lat} at the field value $h_{\star}$ is fixed by \cref{fixedpower}, i.e. $\tilde{\lambda}(h_{\star},t_{{\rm eq}})=\lambda_{\star}$. This, together with the relation defining the boundary
\cref{delta1} forms a system of two equations with four unknowns $\{t_{{\rm eq}},h_{\star},\mathcal{M},h_{{\rm end}}\}$. In order to close the system we add the equation for the number of $e$-folds (if not otherwise specified we use $N_{\star}=60$) and $\epsilon_V(h_{{\rm end}})=1$. Summarizing, we want to solve the following system of equations
\begin{equation} \label{system}
\tilde{\lambda}(h_{\star},t_{{\rm eq}})=\tilde\lambda_{\star},\qquad\mathcal{M}=\frac{\lambda_{{\rm eq}}^{1/4}}{\sqrt{2}y_{{\rm eq}}}m_{t}^{{\rm EW}}e^{t_{{\rm eq}}} ,\qquad 
N_{\star}=\int_{h_{{\rm end}}}^{h_{\star}}\frac{dh}{\sqrt{2\epsilon}}, \qquad \epsilon_V(h_{{\rm end}})=1.
\end{equation}
We used $\phi_{{\rm eq}}=y_{{\rm eq}}^{-1}\,m_{t}^{{\rm EW}}e^{t_{{\rm eq}}}$
to rewrite \cref{delta1} in terms of $\mathcal{M}$ in the second equation.
In practice we do not solve explicitly the last equation but approximate $h_{\star}\gg h_{{\rm end}}\simeq0$  as it turns out that $h_{{\rm end}}$ is always one or two orders of magnitude smaller than $h_{\star}$. The values for $\{\tilde\lambda_{\star},N_{\star},m_{t}^{{\rm EW}}\}$ are fixed. Further, $\lambda_{{\rm eq}},y_{{\rm eq}}$ are the SM running couplings evaluated at $t_{{\rm eq}}$; they depend implicitly on the boundary conditions at the electroweak scale for $\lambda,y$, i.e.  on the mass of the top and the Higgs measured at the LHC.

\subsubsection{Boundary conditions at the electroweak scale}

\begin{figure}[t]
	\centering
	\includegraphics[width=0.47\textwidth]{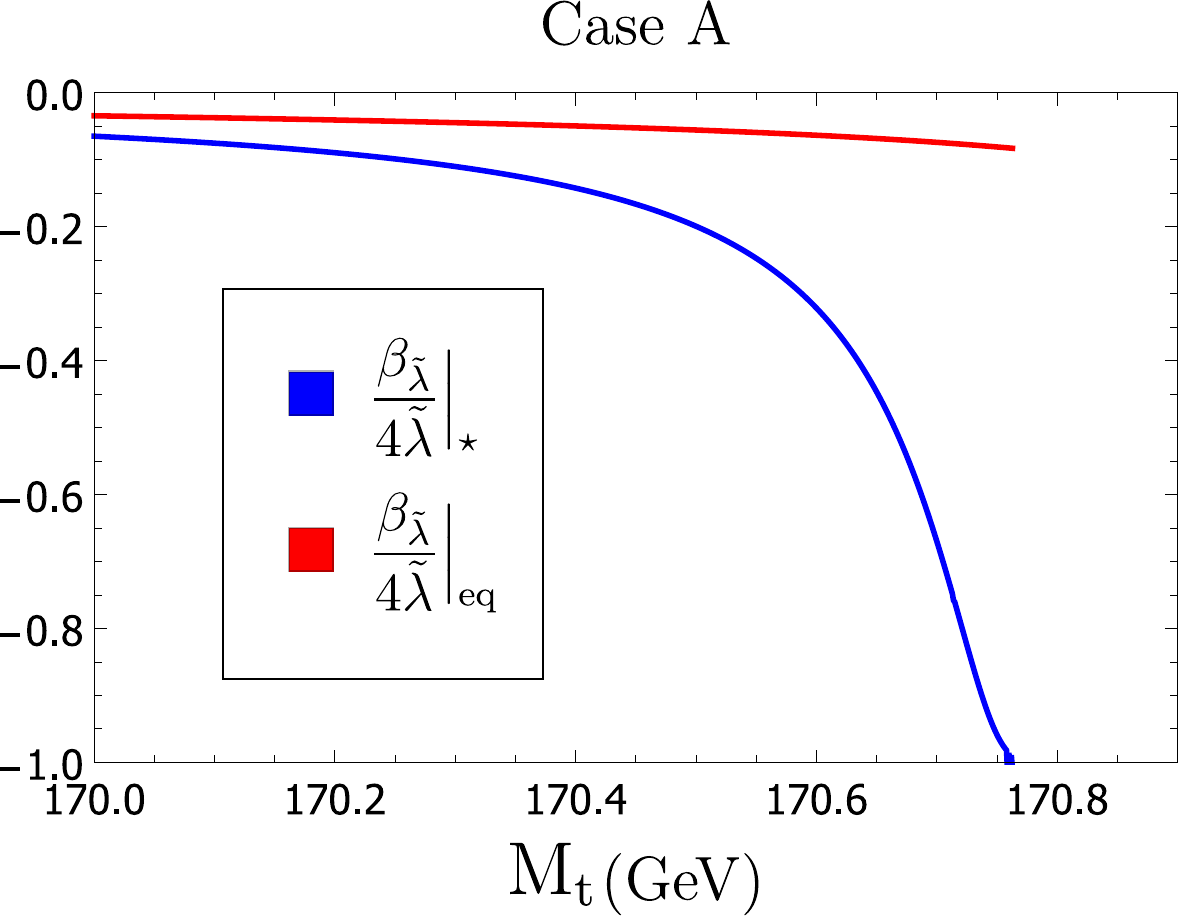} \hspace{0.48cm}
	\includegraphics[width=0.47\textwidth]{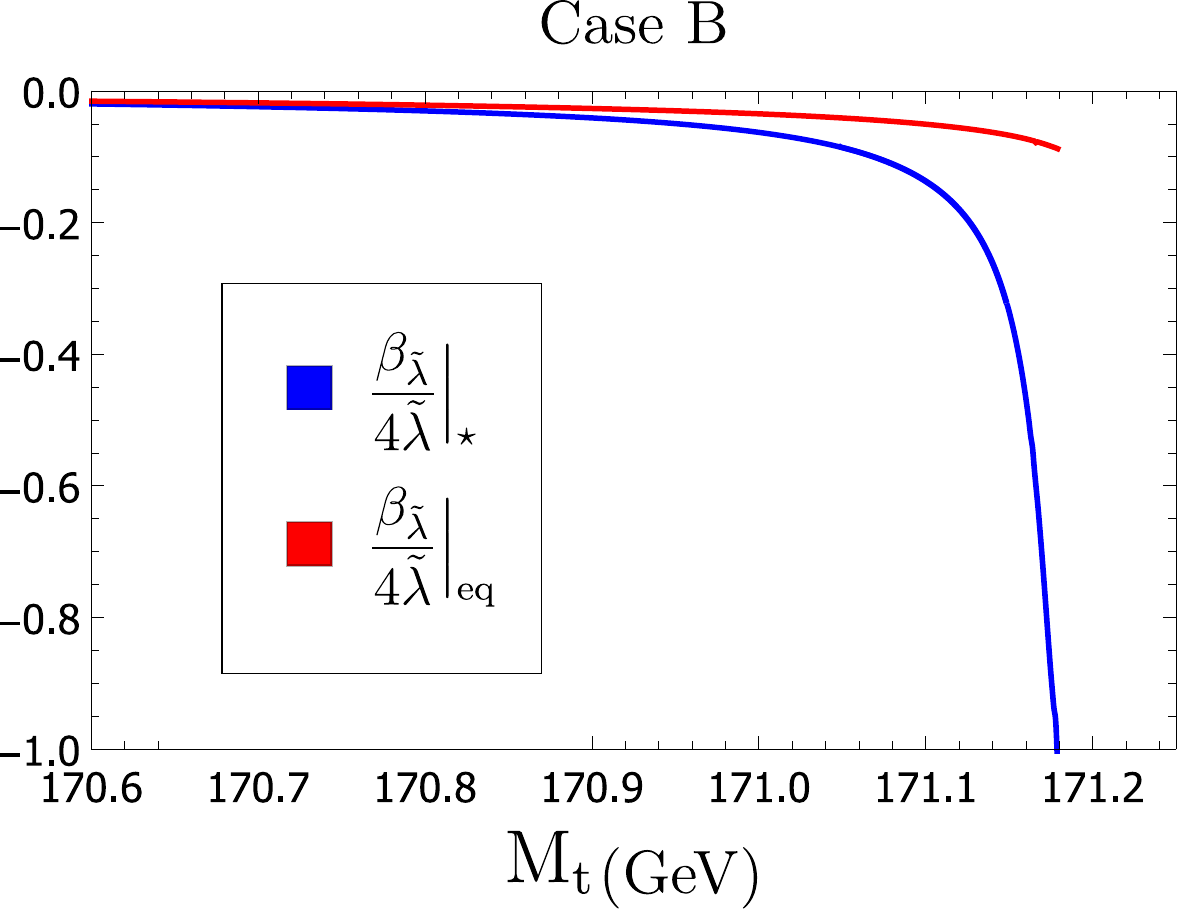} \hspace{0.48cm}
	\caption{ 
		\textit{The value of $(\beta_{\tilde\lambda}/\tilde\lambda)_{\rm{eq}}$ (red curves) depends only on the SM couplings at the EW scale and on the running (no threshold corrections are included). Its value translates in the actual size of the RG correction (blue curves) through the enhancement given in \cref{correction}. 
                  For $(\beta_{\tilde\lambda}/\tilde\lambda)_{\star}<-1$, the potential cannot sustain 60 $e$-folds of inflation anymore. In case B, due to the gauge boson contributions, the absolute value of the correction takes almost a step function shape. This causes a smaller region in parameter space
                  with respect to case A for which the predictions are sensitive to the RG flow.}}
	\label{sizeofbeta}
\end{figure}

\begin{figure}[t]
	\centering
	\begin{subfigure}{1\textwidth}
		\includegraphics[width=0.8\textwidth]{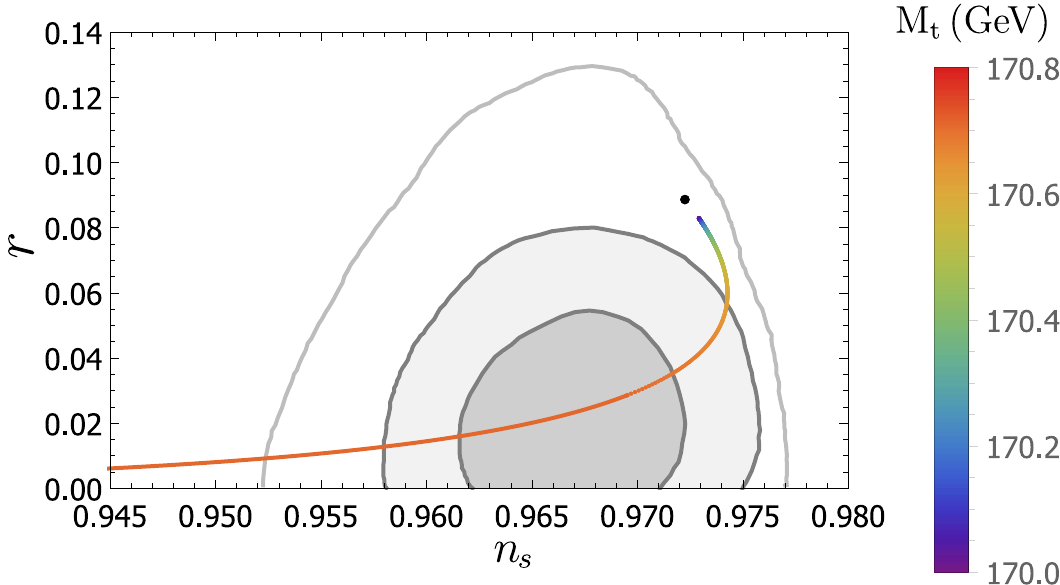}
		\caption{\textit{Inflationary predictions for varying top mass. The black dot represents the tree level result that is reached for unrealistic values $M_{\rm{t}}<160\rm{GeV}$}.}
		\label{planck1}
	\end{subfigure}
	\begin{subfigure}{1\textwidth}
		\includegraphics[width=0.8\textwidth]{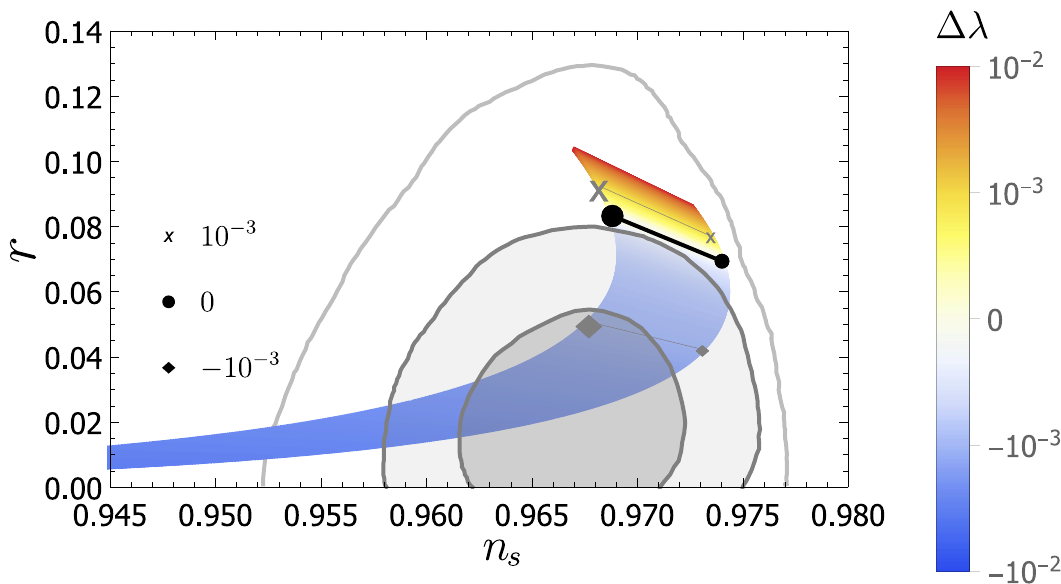}
		\caption{\textit{Inflationary predictions for $M_t=170.5\, \rm{GeV}$ and $M_{h}=125.6\,\rm{GeV}$ fixed and varying threshold conditions.  The cross, circle, and diamand markers correspond to $\Delta \lambda = 10^{-3},\, 0$ and $10^3$ respectively. Results are shown for the range $N_*=50$ (big marker) to $N_*=60$ (small marker). Note that, in agreement with \cref{texplanation}, negative kicks lead to a bigger spread in the predictions.}}
		\label{planck2}
	\end{subfigure}
	\caption{\textit{Case A: predictions for the spectral index $n_s$ and tensor-to-scalar ratio $r$ compared to the 2018 ($1$ and $2\,\sigma$ in the grey shaded regions) and the 2015 Planck data (the external light grey line is the $2\,\sigma$ contour). The effect of threshold corrections is completely degenerate with changing EW boundary conditions.}}
\end{figure}

First we discuss the results without threshold corrections, when the observables only depend on the boundary conditions at the EW scale.  Without running, $\mathcal{M}$ would be fixed by $\tilde{\lambda}_{\star}$ in \cref{fixedpower} via \cref{rescale}, and its constant value would define the matching point \cref{delta1}.  Instead $\{\mathcal{M},t_{{\rm eq}}\}$ are coupled by the first two equations in \eref{system}. Before solving it numerically, it is useful to build some idea about what results to expect. For this we solve $\mathcal{M}$ from the first equation by neglecting the running of $\tilde{\lambda}$ in the large field regime, i.e the second term in \cref{lat}; this will give some numerical correction but does not change the qualitative nature of the solution. We find $\sqrt{6}\mathcal{M}\simeq\tilde{\lambda}_{\star}^{3/8} \lambda_{{\rm eq}}^{-1/8}$. Substituting in the second equation, and solving for $\lambda_{\rm eq}$ gives
\begin{equation}
\lambda_{\rm eq} = \lambda_{\star}
\( \frac{ y_{\rm eq}}{\sqrt{3}m_t^{\rm EW} e^{t_{\rm eq}}} \)^{8/3} e^{-8t_{\rm eq}/3} \equiv
\alpha e^{-8t_{\rm eq}/3}\, .
\label{lambdacurve}
\end{equation}
Thus $\lambda_{{\rm eq}}$ is given by the intersection of the two curves
\be
\lambda(t)=\alpha e^{-8t/3},
\label{intersection}
\ee
with $\alpha$ depending (weakly) on the boundary conditions at the EW scale.  For example, for fixed Higgs mass a larger top mass will give a larger matching point $t_{\rm eq}$, and thus a smaller $\lambda_{\rm eq}$ as the coupling value decreases with renormalization time $t$ (see \cref{texplanation}).  To first approximation we have that the corrections to the inflationary parameters go as $(\beta_{\tilde{\lambda}}/{4\tilde{\lambda}})|_{\star}$. From \cref{betatilde} and \cref{lat} we arrive at
\begin{equation}
\label{correction}
\left(\frac{\beta_{\tilde{\lambda}}}{4\tilde{\lambda}}\right)\Big|_{\star}=\left(\frac{\beta_{\tilde{\lambda}}}{4\tilde{\lambda}}\right)\Big|_{\rm{eq}}\left(1+\left(\frac{\beta_{\tilde{\lambda}}}{4\tilde{\lambda}}\right)\Big|_{\rm{eq}}\ln(...)\right)^{-1}, \qquad
\left(\frac{\beta_{\tilde{\lambda}}}{4\tilde{\lambda}}\right)\Big|_{\rm{eq}}=-\frac{3y_{{\rm eq}}^{4}}{32\pi^{2}}\frac{1}{\lambda_{{\rm eq}}}\propto\frac{1}{\lambda_{{\rm eq}}}.
\vspace*{0.1cm}
\end{equation}
It follows that the corrections to the observables parameterized by $\big|\beta_{\tilde \lambda}/(4\tilde \lambda)\big|_\star$ increase for smaller $\lambda_{\rm eq}$, i.e. for a larger top mass. This is illustrated in the left plot of \cref{texplanation}.  In \cref{correction} $\ln(..)$ is the log appearing in \cref{lat}, which numerically is order $\mathcal{O}(10)$ for different boundary conditions. The log enhances the size of the corrections as is shown in the left plot of \cref{sizeofbeta}. For example, for boundary conditions such that $\beta_{\tilde \lambda}/(4\tilde \lambda )\big|_{\rm eq} = 10^{-2}-10^{-1}$, the corrections are already order one.  For larger corrections it is no longer possible to obtain $N_\star=60$ $e$-folds of inflation. This can also be seen from \cref{fig1}: the larger the top mass the larger the correction, but we see that this also pushes the maximum of the potential to smaller Higgs field values until the region on the left is too small to support 60 $e$-folds.

The full numerical results for the observables $n_s$ and $r$ are shown in \cref{planck1} for different top masses.  We also plotted the contours of the 2015 and 2018 Planck data \cite{Planck,Akrami:2018odb}.  Although the tree-level results are outside the $2\sigma$-contours of the latest Planck data, the running corrections can bring the model back into the region favored by Planck.

\subsubsection{Threshold corrections}
Let us now include threshold corrections, which we model by a shift in $\lambda\rightarrow\lambda+\Delta\lambda$ at the boundary between the small and large field regime. $\Delta\lambda$ has to be considered as the sum of the contributions from higher dimensional operators to the running of $\lambda$ \cite{cliffnew,Fumagalli,shap,critical1,Enckell:2016xse,Bezrukov:2017dyv,Fumagalli:2017cdo}. This means that at $t_{{\rm eq}}$, the value of $\lambda$ that is matched to the tilde parameters is shifted by
\begin{equation}
\lambda_{{\rm eq}}^{{\rm new}}=\lambda(t_{{\rm eq}})+\Delta\lambda
\end{equation}
If we assume as in \cite{Shaposhnikov:2020fdv} that new physics implies a shift in the beta function of $\lambda$ on the order of $\delta\beta_{\lambda}\sim 1/(4\pi)^2$ at the matching scale, the effective shift seen at the inflationary scale by $\lambda$ would be given approximately by $\Delta \lambda \sim \delta\beta_{\lambda} \ln (\phi_*/\phi_{\mathrm{eq}})\sim 10^{-2}$.  In principle there is also a jump in the Yukawa coupling, which we ignore (this gives a degeneracy with the EW boundary value of the Yukawa coupling) as the relative correction is small.  To see the effect of the threshold corrections, we solve the same system of equations \eref{system}, but with the substitution $\lambda_{{\rm eq}}\rightarrow\lambda_{{\rm eq}}^{{\rm new}}$.  We fix the boundary conditions at the EW scale and let $\Delta\lambda$ vary. In order to understand the numerical results we can go through the same steps as before, with the only difference that \eref{intersection} now becomes $\lambda(t)+\Delta\lambda=\alpha e^{-8t/3\,}$. For fixed EW boundary conditions $t_{{\rm eq}}$ is now given by the intersection between the shifted curve and the same $\alpha e^{-8t/3\,}$ as before.  It follows that a positive/negative $\Delta\lambda$ will cause a smaller/larger correction that goes  ``up/down" in the $(n_{s},r)$-plot compared to the tree level result, as illustrated in \cref{texplanation}.

The predictions for $n_s$ and $r$ in presence of threshold correction are shown in \cref{planck2}, where we now also showed the band range for $N_\star = 50-60$. The curve trajectory in the $(n_{s},r)$-plot obtained by decreasing the shift $\Delta \lambda$ is fully degenerate with the curve obtained by increasing the top mass, as can be seen by comparing with \cref{planck1}.  This can easily be understood by looking at the systems of equations solved, and can also be clearly seen from \cref{texplanation}:  increasing the top mass and decreasing the shift $\Delta \lambda$ both have the same effect of decreasing $\lambda_{\rm eq}$, and thus increasing the RG corrections to inflation.

\subsection{Matching and running: other cases}

\begin{figure}[t]
	\centering
	\includegraphics[width=0.90\textwidth]{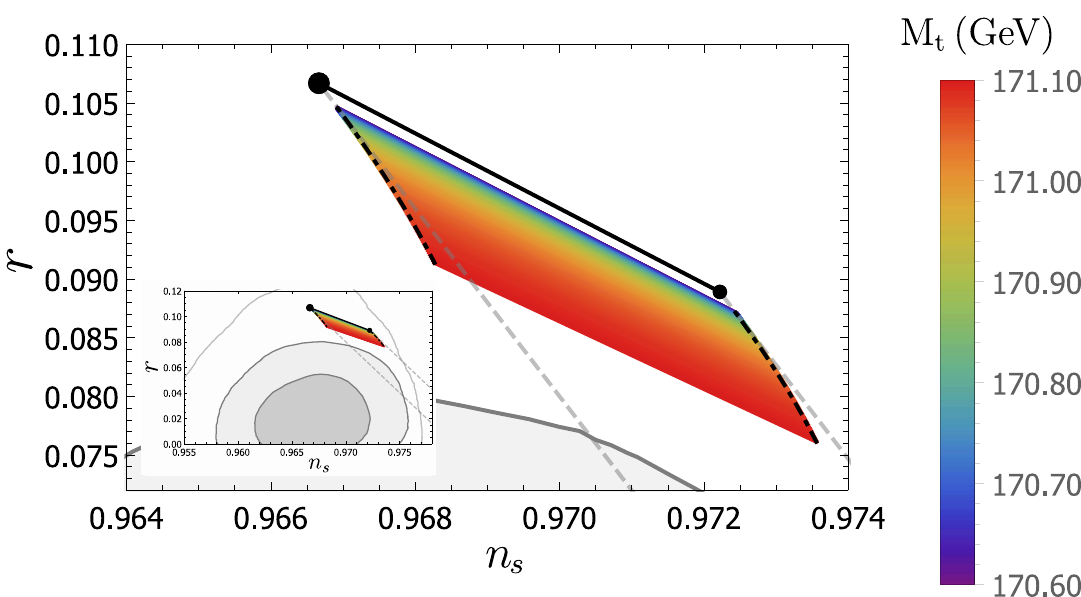} \hspace{0.5cm}
	\caption{ 
		\textit{$(n_s,r)$ predictions for Case B without threshold corrections and for $\alpha_0=1$. The dashed line represents the analytic approximation of eq. \eqref{observables}. We have zoomed in on the Planck plot since the inflationary observables are less RG dependent than in case A. This outcome can be explained (see main text) by looking at fig. \ref{sizeofbeta}.}}
	\label{caseb}
\end{figure}

In case B the gauge bosons are back in the spectrum during inflation. We proceed in  the same way as in the previous section, but with different RGEs in the large field regime \cref{inflationRGE1} with $(f_1,f_2)=(1,1)$. The running of $\tilde \lambda$ depends on the gauge couplings only through the combinations
$g^2_i/\tilde{\alpha_0}$, and thus we add to \cref{matchingcondition} the following matching condition for the gauge couplings
\begin{equation}
\tilde{g}^2_{i\rm{eq}} \equiv \frac{g^2_{i\rm{eq}}}{\tilde{\alpha}_{0\rm{eq}}}=g^2_{i\rm{eq}}\lambda_{\rm{eq}}^{-1/3}\alpha^{-1}_0\mathcal{M}^{4/3}.
\end{equation}
The results of the numerical implementation for case B (without threshold corrections) are given in \cref{caseb}. As in case A the tree level results are never reached for realistic values of the top mass. However, for the values of the boundary conditions that allow inflation to happen,  there is less dependence on the RG flow. This can be explained by looking at the right-hand plot in \cref{sizeofbeta}. The positive contribution from the gauge bosons to $\beta_{\tilde \lambda}$ leads to $(\beta_{\tilde \lambda}/\tilde{\lambda})_{\rm {eq}}$ changing rapidly over a relatively narrow range of top masses of ${\cal O}(0.1)\,$GeV, roughly between $171-171.2\,\rm{GeV}$. Since the corrections proportional to $(\beta_{\lambda}/\lambda)_{\star}$ are enhanced through \cref{correction}, we jump quickly from having zero corrections to spoiling inflation completely, i.e. from $(\beta_{\lambda}/\lambda)_{\star}\simeq 0$ to $(\beta_{\lambda}/\lambda)_{\star}\simeq -1$.  This also means that the inflationary predictions cannot be brought back in the $2\sigma$-contour of the latest Planck data by the RG flow corrections.  Since changing EW boundary conditions and varying matching corrections is degenerate, this result is robust.

This is in contrast with case A,  as can be seen from the left-hand plot in \cref{sizeofbeta}.  In this case the corrections grow much more gradually with changing top mass,  and change from percent level to order one over a much wider range of top masses of ${\cal O}(1)\,$GeV.  As a result the inflationary predictions vary over a wider range as well, and as noted before, in case A it is possible that the RG corrections bring the predictions back in the region favored by Planck.

In case C the implementation is very similar to the previous cases with the quantitative difference that now, since the top quark decouples, $\beta_{\tilde\lambda}>0$ in the large field regime. This implies positive corrections to $n_s$ and $r$, as can be understood from the approximate solution \cref{observables}. Thus, by increasing the top mass (i.e. increasing the size of the corrections), the inflationary parameters ``move up"  in the Planck plot to higher values of the tensor-to-scalar ratio $r$ compared to the tree-level result, farther outside region favored by Planck.
Finally, case D gives trivially the tree level results since $\beta_{\tilde \lambda}=\mathcal{O}(\delta^{-1})$.

\section{Conclusions}\label{outlook}

We studied the effects of the renormalization group flow on inflationary models that are embedded in low energy (beyond the) Standard Model theories. The inflationary predictions depend on the renormalization group equations during the inflationary stage and on a new kind of UV sensitivity \cite{cliffnew} present in inflationary models with non-renormalizable operators.

In this paper we developed the necessary tools to analyse the effects of the running couplings, and of the threshold corrections parameterizing the UV physics, on the inflationary predicitons. Results have been applied to the SM with non-derivative couplings to gravity and different cases have been considered in which apart from the Higgs also gauge boson/fermions are non-minimally coupled to gravity.

We have calculated the one-loop beta functions in the large field regime using a covariant approach that takes into account the non-trivial geometry of the field space manifold. The results for the RGEs are summarized in \cref{inflationRGE1}. Given the simple form of the beta functions, it may seem that by introducing the covariant formalism we used a sledgehammer to crack a nut. However, as we saw for example in case B, the covariant formalism is necessary to have a consistency check and thus obtain reliable results. This suggests that in general a covariant formalism is desirable to compute the RGEs in presence of a non-flat metric in field space.

The explicit dependence of $n_s$ and $r$ on the running coupling $\tilde \lambda$ defined in \cref{rescale} can be computed via an analytical approximation with results given in \cref{observables}. In inflationary models in which, in the large field limit, the potential asymptotes to a constant value exponentially fast as a function of the canonical field, the RG corrections to $n_s$ and $r$ disappear at first order in the $1/N_\star$ expansion due to a cancellation between the running dependence of the slow-roll parameters and of the number of $e$-folds. This is the case for the large class of cosmological attractors that includes Higgs inflation as a particular case \cite{Fumagalli,Fumagalli:2016sof}. As \cref{observables} shows, such an insensitivity to running effects does not happen in general, and not in models where the approach to a constant is polynomial.

The full non-linear RG dependence for the original new Higgs inflation proposal with only a non-minimal coupling for the Higgs field (case A) is shown in figure \ref{planck1} (as a function of the boundary conditions at the EW scale), and in figure \ref{planck2} (as a function of the threshold corrections). In figure \ref{caseb} the sensitivity of the predictions when the gauge bosons are also non-minimally coupled (case B) is illustrated.  In both cases the tree level results are modified by the running corrections.  In case A this allows to `push back' the prediction into the region favored by the Planck data, whereas in case B the running corrections ruin the flatness of the inflationary potential before the 2$\sigma$-region is reached. 

The general lesson is that in these type of models the classical description is not enough to sensibly compare the predictions to the Planck data.

\section*{Acknowledgements}
We thank Cristiano Germani and Sander Mooij for helpful discussions. JF is supported by the European Research Council under the European Union’s Horizon 2020 research and innovation programme (grant agreement No 758792, project GEODESI).  When this work started MP and JF were supported by the Projectruimte “Keeping track of time during inflation” grant from the Dutch Organization for Scientific Research (NWO).

\appendix

\section{Calculation of the self-energies}
\label{A:details}

In this appendix we provide more details for the calculation of the self-energies of the Higgs, Goldstone, fermion and gauge fields.  We first determine the self-energies in the Abelian-Higgs model augmented with a single fermion.  This not only simplifies the calculation of the two-point funtions but also avoids much notational clutter. Sections \cref{A:renormalized_action,feynmannappendix,A:couplings} give the expressions for the $U(1)$ theory. In \cref{A:abel} we calculate the two-point funtions for the Higgs, gauge and fermion field in the Abelian model. These results for the self-energies are subsequently generalized to the full SM gauge group in \cref{A:SM}.

\subsection{Renormalized action}
\label{A:renormalized_action}

We give the Lagrangian, counterterms and Feynman rules for the Abelian-Higgs model with a single Dirac fermion field.  Rescaling the bare parameters \cref{Zelement} the renormalized action is 
\begin{align}
\mathcal{L}&=
-\frac12 K_{\phi}(\varphi^a)  Z_\phi\partial_\mu \varphi^a \partial^\mu \varphi_a
- \frac14 K_{A}(\varphi^a)  Z_A (F_{\mu\nu})^2
+ K_{\psi}(\varphi^a)   Z_{\psi} \bar \psi (i\slashed{\partial}) \psi
\nn \\
& \hspace{0.43cm}-   Z_\phi^2 Z_\lambda V(\varphi^a) -
         Z_\psi Z_y Z_\phi^{1/2}\frac{y}{\sqrt{2}} \bar \psi F_{\psi}(\varphi^a) \psi
         - Z^2 Z_{A}   Z_\phi \frac12   g^2 A^2 F_A(\varphi^a)
\nn\\
& \hspace{.43cm}
+Z_g Z_A^{1/2} Z_\psi \(g q_L \bar \psi \slashed{A} P_L \psi +g q_R \bar \psi \slashed{A} P_R
\psi\)
+Z_g Z_A^{1/2} Z_\phi^{1/2}  K_{\phi}(\varphi^a)  g q_H A  (\varphi \partial \theta - \theta \partial \varphi).
\label{Lexp}               
\end{align}
With the charge of the Higgs field fixed to unity, gauge invariance implies the relation $q_L -q_R+q_H$ for the charges of the left and right handed fermions and the Higgs field. We can then identify the $U(1)$ symmetry with hypercharge by equating $2Y_i = q_i$.
Further, 
\begin{align}
V &=\frac{\lambda}{4} (\varphi^2 + \theta^2)^2, &
F_{\psi} &= \( \varphi-i\gamma^5\theta\), &
F_{A} &=K_{\phi} (\varphi^2 + \theta^2), &
\nn\\
K_{\phi} &= \(1+ Z_\delta \delta \), &
K_{\psi} &= \(1+ Z_{\alpha_F}Z_\delta \alpha_F \delta\),&
K_{A} &= \(1+  Z_{\alpha_A}Z_\delta \alpha_A \delta\)
\label{VF}
\end{align}
with $\delta$ given in \eref{delta}, and we introduced the notation
$Z_\delta = {Z_\phi^2 Z_\lambda}/{Z_\M^4} $. 
The gauge fixing term for the generalized $R_\xi$-gauge is
\begin{align}
\mathcal{L}_{\rm GF} &= -\frac1{2\xi} K_{A}(\phi)\( Z_A^{1/2}\partial^\mu A_\mu - Z_g Z_\phi g \xi \frac{K_{\phi}(\phi)\phi}{K_{A}(\phi)} \theta\)^2
\label{L_GF}
\end{align}
The  $A_\mu \partial^\mu \theta$-term cancels the interaction in the Higgs kinetic terms, and the quadratic terms are diagonal.  We work in Landau gauge $\xi=0$, for which the ghosts fields decouple. %

\subsection{Feynman rules}\label{feynmannappendix}

In this subsection we will give the Feynman rules for the action in terms of the covariant fields.   We expand the action \cref{Lexp,L_GF} in covariant fields using \cref{covnoncov,taylor}, where we use the notation in \cref{T:notation}.
This gives the interaction Lagrangian which the defines the various couplings
\begin{equation}
\begin{split}
  \mathcal{L}_{\rm int} = & -\lambda_{m Q^{\phi} n Q^{\theta}} (Q^{\phi})^m (Q^{\theta})^n - y_{m Q^{\phi} n Q^{\theta}} (Q^{\phi})^m (Q^{\theta})^n \bar \psi (i\gamma^5)^n \psi \\&-y_{2\psi} \bar \psi \psi +g_L \bar \psi Q^{A} P_L \psi + g_R \bar \psi Q^{\phi} P_R \psi
  \nn \\
  &-(g_{ Q^A \partial Q^{\phi} m Q^{\phi} n Q^{\theta} } \partial Q^{\phi} - g_{ Q^A \partial
	Q^{\theta} m Q^{\phi} n Q^{\theta}} \partial Q^{\theta} ) Q^A (Q^{\phi})^m (Q^{\theta})^n
\\&-g_{2Q^A mQ^{\phi} nQ^{\theta}} (Q^A)^2 (Q^{\phi})^m (Q^{\theta})^n+...
\end{split}
\label{vertices1}
\end{equation}
Equivalently we expand the kinetic terms, for example \be \mathcal{L}_{\rm{k}} =- \mathcal{K}_{Q^{\phi} 2 \partial Q^I} (Q^{\phi})^2 (\partial Q^I)^2 - \mathcal{K}_{2Q^{\theta} 2 \partial Q^I} (Q^{\theta})^2 (\partial Q^I)^2 +...  \ee All interactions are defined with a minus sign (the only exception is for one of the derivative interactions and the fermion-gauge interaction), and without numerical factors. This means that for a vertex with $m$ $Q^{\phi}$-fields and $n$ $Q^{\theta}$-fields and with or without fermion/gauge lines we have, respectively:
\begin{align}
V^{(m Q^{\phi}nQ^{\theta})}&= (-i) m! n!  \lambda_{m Q^{\phi}n Q^{\theta}},\nn\\
V^{(m Q^{\phi}nQ^{\theta}2\psi)}&= (-i)m! n!  y_{m Q^{\phi}
	nQ^{\theta}}(i\gamma^5)^n,\nn \\
V^{(m Q^{\phi}nQ^{\theta}2Q^{A})}&= (-i) 2! m! n!  g_{2Q^{A}m Q^{\phi}nQ^{\theta}}.
\label{vertices2}
\end{align}
For the derivative interaction we get
\be
\begin{split}
	&V^{(Q^A \partial Q^{j} m Q^{\phi} nQ^{\theta})} =  -i g_{ Q^A \partial Q^{j}  m Q^{\phi} n Q^{\theta} } (-i
	p^\mu),\quad j=\{\phi,\theta\}\\
	&V^{2Q^{j} 2 \partial Q^I}\quad\qquad\,\,=2!2!(-i\mathcal{K}_{2Q^{j} 2 \partial Q^I})p^{\mu}p_{\mu},\quad j=\{\phi,\theta\}
\end{split}
\ee
with $p$ the momentum running through the vertex.  The fermion, scalar and gauge propagators are given by:
\begin{align}
-i D_\psi(p) &= K_{\psi}^{-1}\frac{-i(-\slashed{p}+m_\psi)}{p^2+m_\psi^2-i\eps},
\nn \\
-i D_{Q^I}(p) &= K_{\phi}^{-1} \frac{-i}{p^2 + (m^2)_I^I-i\eps},
\nn \\
-i D_{\mu\nu}(p)
&\stackrel{\xi_G = 0}{ =}-i K_{A}^{-1}\frac{g_{\mu\nu}-\frac{p_\mu p_\nu}{p^2}}{p^2 +m_A^2 -i\eps}.
\label{propagators}
\end{align}
The masses are given in \cref{mass1,mass2}.

Finally, we need the countertems in the quadratic action, We define the notation
\be
{\cal L}_{\rm ct} = -\sum_I \( \frac12 Z_{2Q^I} (\partial Q^I)^2+ Z_{c_I 2Q^I} c_{c_I 2 Q^I} (Q^I)^2 \)
+Z_{2\psi_{L,R}} \bar \psi_{L,R} \slashed{\partial} \psi_{L,R} - Z_{y2\psi} y_{2\psi } \bar \psi \psi
\ee
with $I$ running over the bosonic fields and $c_I=\{\lambda,g\}$ labels the coupling that the counterterm normalizes.  The counterterm can be expressed in terms of the ``elementary" ones \cref{Zelement} as
\begin{align}
Z_{2Q^{\phi}}&=Z_{2Q^{\theta}} = Z_\lambda Z_\phi^3 Z_{\mathcal{M}}^{-4},&
Z_{\lambda2Q^{\phi}}&=Z_{\lambda2Q^{\theta}}  = Z_\lambda Z_\phi^2,                                                                                                                    
\nn \\
Z_{2Q^A}&=Z_A Z_{K_A},&
Z_{g2Q^A}& =  Z_{K_A} Z_g^2 Z_A Z_{\alpha_A}^{-1},                               
\nn \\
Z_{2\psi_{L,R}} &=Z_{\psi_{L,R}} Z_{K_\psi},&
Z_{y2\psi}& = Z_\psi Z_y Z_\phi^{1/2},
\label{Z2pnt}
\end{align}
with
\be
Z_{K_A} = Z_{\alpha_{0A}} \(Z_\lambda Z_\phi^2 Z_{\mathcal{M}}^{-4}\)^{1+\frac{n_A}{2}} ,\quad
Z_{K_\psi} = Z_{\alpha_{0f}} \(Z_\lambda Z_\phi^2 Z_{\mathcal{M}}^{-4}\)^{1+\frac{n_f}{2}}.
\ee
%

 \subsection{Coupling strengths}
 \label{A:couplings}
 
 We list the relevant vertices in the theory with a $U(1)$ gauge group for case B, where both fermions and gauge bosons are in the spectrum.
 Case A can be obtained from this result by setting $g \to g K_{A}(\phi)$, and then subsequently integrating out the gauge bosons; thus effectively we can set the gauge coupling to zero. Case C is obtained by rescaling the Yukawa coupling $y \to y/K_{\psi}$, which makes the fermion contribution subleading, and effectively we can set the Yukawa to zero. Case D is obtained by rescaling both $g$ and $y$ with the appropriate metric factors; the net effect is that both fields decouple and we can effectively set both couplings to zero. All couplings are evaluated on the background and we use the notation that the first term in the curly brackets is the leading result in the SM limit ($\delta \ll 1$), and the second term the leading result in the inflationary regime ($\delta \gg 1$).

 The vertices derived from the potential are valid in all four cases. For example, $4! \lambda_{4Q^{\phi}} = \nabla_{\phi}^{4}V $ and $3!\lambda_{Q^{\phi}2Q^{\theta}} = \nabla_{(\theta}\nabla_{\theta}\nabla_{\phi)}V$. The full results are given by
\begin{align}
\lambda_{2Q^{\phi}} & =\frac{\lambda \phi^2}{2} \{3,1\}, &
	\lambda_{2Q^{\theta}} &=\frac{\lambda \phi^2}{2} \{1,3\},&
	\lambda_{3Q^{\phi}}&=\frac{\lambda\phi}{3}\{3,-1\},&               \lambda_{4Q^{\phi}} & = \frac{\lambda}{12}\{3, 5  \}, \nn \\
	\lambda_{4Q^{\theta}} &= \frac{\lambda}{4}\{1, -9 \},&
	\lambda_{Q^{\phi}2Q^{\theta}} & = \lambda\phi\{1, -3 \},&
	\lambda_{2Q^{\phi}2Q^{\theta}} & = \frac{\lambda}{2}\{1, 15 \}. 
\end{align}
The vertices derived from the Yukawa interactions are $y_{Q^{\phi}} =  \nabla_{\phi}F_{\psi}$, $y_{Q^{\theta}} = i\gamma_{5} \nabla_{\theta}F_{\psi}$ etc.
\begin{align}
  y_{Q^{\phi}} &= \frac{y}{\sqrt{2}}\{1,1\},&
y_{Q^{\theta}} &= \frac{y}{\sqrt{2}}\{1,1\},&
y_{2Q^{\phi}} & =-\frac{y}{\sqrt{2}\phi} \{0,1\},&
  y_{2Q^{\theta}} & = -\frac{y}{\sqrt{2}\phi} \{0,1\}.
\end{align}
The vertices involving gauge fields are
\begin{align}
g_{2Q^A2Q^{\phi}} &= \frac{g^{2}}{6}\{ 3, -5\delta \},&
g_{2Q^A2Q^{\phi}} &= \frac{g^{2}}{2}\{ 1, 5\delta \},&
g_{2Q^AQ^{\phi}} &= g^{2}\phi\{1, \delta\}, \nn \\
g_{Q^A\partial Q^{\phi}Q^{\theta}} &= g\{1, \delta \},&
g_{A \bar \psi_{s} \psi_s} &= g q_s\{1,1\}.
\end{align}
with $s=L,R$.
The relevant couplings coming from the kinetic terms are 
\begin{equation}
  \mathcal{K}_{2\partial Q^A2Q^{\phi}} = \frac{\alpha_{A0}\delta}{3\phi^2}\{0, -1 \},\quad \mathcal{K}_{2\partial Q^A 2Q^{\theta}} = \frac{\alpha_{A0}\delta}{\phi^{2}} \{0, 1\}
  \label{K_int}
\end{equation}

\subsection{Two-point functions in the Abelian-Higgs model}
\label{A:abel}

With the Feynman rules in hand we can calculate the one-loop self-energies of the Higgs, Goldstone, fermion, and $U(1)$ gauge boson field. We only include the interactions from the kinetic terms \cref{K_int} that are not suppressed by $\delta^{-1}$. The results are
\begin{align}
  \Pi_{Q^{\phi}}\left(p^{2}\right) &= \frac{1}{8\pi^{2}\epsilon}\bigg\{
\big(18\lambda_{3Q^{\phi}}^{2}K_\phi^{-1}+ 12\lambda_{4Q^{\phi}}m_{h}^{2}\big)K_\phi^{-1} + \big(2\lambda^{2}_{Q^{\phi}2Q^{\theta}}K_\theta^{-1} + 2\lambda_{2Q^{\phi}2Q^{\theta}}m_{\theta}^{2}\big)K_\theta^{-1} \nn \\
&+\big( 6g_{2Q^{A}2Q^{\phi}}m_{A}^{2} + 6g_{2Q^{A}Q^{\phi}}^{2} + \frac{3}{4}K_\theta^{-1}\left(g_{Q^{A}\partial Q^{\phi}Q^{\theta}} + g_{Q^A\partial Q^{\theta} Q^{\phi}} \right)^{2}p^{2} -6  \mathcal{K}_{2Q^{\phi} 2 \partial Q^{A}} m_A^4\big)  K_A^{-1}\nn \\
  &- \big(2y_{Q^{\phi}}^{2}( p^{2} + 6m_{\psi}^{2} )K_{\psi}^{-1} + 8y_{2Q^{\phi}}m_{\psi}^{3}\big)K_{\psi}^{-1} \bigg\}  - \Big[ \left( Z_{\partial Q^{\phi}} -1 \right)p^{2}K_\phi+ \left( Z_{\lambda 2Q^{\phi}} - 1 \right)2\lambda_{2Q^{\phi}} \Big],
\label{Pi_h}
\\
\Pi_{Q^{\theta}}\left(p^{2}\right) &= \frac{1}{8\pi^{2}\epsilon}\bigg\{ 
\big(12\lambda_{4Q^{\theta}}m_{\theta}^{2} + 2\lambda_{2Q^{\phi}2Q^{\theta}}m_{h}^{2} + 4K_\phi^{-1}\lambda^{2}_{Q^{\phi}2Q^{\theta}} \big)K_\theta^{-1}\nn \\
&+ \big(6g_{2Q^A2Q^{\theta}}m_{A}^{2} + \frac{3}{4}K_\phi^{-1}\left(g_{Q^A\partial Q^{\phi}Q^{\theta}} + g_{Q^A\partial Q^{\theta} Q^{\phi}} \right)^{2}p^{2} -6  \mathcal{K}_{2Q^{\phi} 2 \partial Q^{A}}  m_A^4 \big) K_A^{-1}\nn \\
&- \big( 2y_{Q^{\theta}}^{2}\left( p^{2} + 2m_{\psi}^{2} \right)K_{\psi}^{-1} - 8y_{2Q^{\theta}}m_{\psi}^{3}\big) K_{\psi}^{-1}
\bigg\}-\Big[ \left( Z_{\partial Q^{\theta}} -1 \right)p^{2}K_\theta^{-1} + \left( Z_{\lambda{2Q^{\theta}}} - 1 \right)2\lambda_{2Q^{\theta}} \Big].
\label{Pi_theta}
\end{align}
For the fermions
\begin{equation}
\begin{split}
  \Pi_{\psi}\left(\slashed{p}\right) &= \frac{1}{8\pi^{2}\epsilon}\bigg\{
  \big( y_{Q^{\phi}}^{2}K_\phi^{-1}( m_{\psi} - \frac{1}{2}\slashed{p}) - y_{Q^{\theta}}^{2}K_\theta^{-1}( m_{\psi} + \frac{1}{2}\slashed{p})
  \big) K_{\psi}^{-1}\\
&\qquad\qquad+ y_{2Q^{\phi}}K_\phi^{-1}m_{h}^{2} - y_{2Q^{\theta}}K_\theta^{-1}m_{\theta}^{2} - 3m_{\psi}K_A^{-1}g^{2}q_{L}q_{R} \bigg\}\\
&\quad- \[\left(  Z_{y{2\psi}} - 1 \right)m_{\psi} + K_{\psi}\slashed{p} P_L\left( Z_{2\psi_L} - 1\right)+ K_{\psi}\slashed{p} P_R\left( Z_{2\psi_R} - 1\right)\].
\label{Pi_psi}
\end{split}
\end{equation}
For the gauge fields
\begin{align}
\Pi^{A}_{\mu\nu} &= \frac{1}{8\pi^{2}\epsilon}\bigg\{ ( 3K_\phi^{-1}K_A^{-1}g^{2}_{Q^{\phi}2Q^A} + 2K_\phi^{-1}g_{2Q^{\phi}2Q^A}m_{h}^{2} + 2K_\theta^{-1}g_{2Q^{\theta}2Q^A}m_{\theta}^{2})g_{\mu\nu}\nn \\
&- K_\phi^{-1}K_\theta^{-1}\left[ \frac{1}{4}\left( g_{Q^A\partial Q^{\phi}Q^{\theta}} + g_{Q^A\partial Q^{\theta} Q^{\phi}} \right)^{2}\left( \frac{p^{2}}{3} + m_{h}^{2} + m_{\theta}^{2} \right)g_{\mu\nu} \right. \nn\\
&\left.-\left(g^{2}_{Q^A\partial Q^{\phi}Q^{\theta}} - g_{Q^A\partial Q^{\phi}Q^{\theta}}g_{Q^A\partial Q^{\theta} Q^{\phi}} + g^{2}_{Q^A\partial Q^{\theta} Q^{\phi}} \right)\frac{1}{3}p_{\mu}p_{\nu} \right] \nn\\
&- \big(\frac{2}{3}\left( p^{2}g_{\mu\nu} - p_{\mu}p_{\nu} \right)g^{2}\left( q_{R}^{2} + q_{L}^{2}\right) +2g^{2}\left(q_{L} - q_{R} \right)^{2}m_{\psi}^{2}g_{\mu\nu}\big) K_{\psi}^{-2}\bigg\}\nn\\
                 &- \left(Z_{2Q^A} - 1\right)K_A^{-1}\left(p^{2}g_{\mu\nu} - p_{\mu}p_{\nu} \right) - \left( Z_{m_{A}^{2}} - 1 \right)g_{\mu\nu}m_{A}^{2}.
\label{Pi_A}                   
\end{align}

\subsection{Two-point functions in the SM}
\label{A:SM}

In this subsection we generalize the $U(1)$ result to the full SM gauge $SU(3)\times SU(2) \times U(1)$. We only include the effects of the top quark yukawa, but neglect the smaller yukawa couplings of the other fermions. To set the notation and our conventions, the covariant derivatives are 
\begin{align}
  D_\mu {\cal H} & = (\partial_\mu - i g a^a_\mu \tau^a - i Y_{\cal H}  g' B_\mu) {\cal H} \nn \\
  D_\mu Q_L& = (\partial_\mu - i g_s f^a_\mu t^a - i g a^a_\mu \tau^a - i Y_Q g' B_\mu) Q_L \nn \\
  D_\mu u_R & = (\partial_\mu - i g_s f^a_\mu t^a - i Y_u g' B_\mu) u_R
 \end{align}              
 with $\{B,a,f\}$ the $U(1)$, $SU(2)$ and $SU(3)$ gauge fields respectively with corresponding gauge couplings $\{g',g,g_s\}$. The hypercharge assignments are $Y_{\cal H} =1/2,\, Y_Q=1/6$ and $Y_u=2/3$. 

\paragraph{Higgs self energy: $\Pi_{h}$.}

The Higgs kinetic term can be written in the form
 \be
 {\cal L}_{\cal H} \supset - K_\phi (D_\mu {\cal H})^\dagger  (D^\mu {\cal H})
 = - K_\phi\bigg[ (\partial \delta \phi)^2 + \sum _{i=1}^3 \Big( (\partial \theta_i)^2 -2g_i  A^i (\phi \partial \theta_i - \theta_i \partial \phi) + g_i^2 \phi^2  A_i^2\Big) + ...\bigg].
 \label{L_Z}
\ee
The gauge boson mass eigenstates are $ A^i = \{A^1,A^2,Z,A_\gamma\}$ with $W^\pm =A^1\pm i A^2,\, Z,A_\gamma$ the usual $W$, $Z$ and photon fields, and the gauge couplings are  $g_i = \frac12 \times \{g,g,\sqrt{g^2 +g^{'2}},0\}$.
The Higgs/Goldstone-gauge couplings for the electroweak mass eigenstates in \cref{L_Z} are exactly of the form of three massive (and one massless $U(1)$ gauge bosons) that have eaten the Goldstone bosons $\theta_i$.  We can choose a gauge fixing term as in \cref{L_GF} but now for each pair of $\{A^i,\theta^i\}$.  Hence, in all loops containing gauge- Higgs interactions we can use a U(1) model and sum over the three gauge couplings \cite{damien2}. In addition, in loops with gauge-Higgs or gauge-Goldstone interactions, we sum over the $n_{\rm GB}=3$ three Goldstone bosons. However, However, in all cases (A-D) the Goldstone fluctuations decouple, and this change will not affect the inflationary RGEs.

The difference in the Higgs-fermion Yukawa coupling compared to the $U(1)$ case is that the top quark is now an $SU(3)$ triplet. We have to sum over the dimension of the representation which picks up a color factor $N_c =3$.

The SM Higgs self energy thus becomes:
\begin{align}
 & \Pi_{Q^{\phi}}\left(p^{2}\right) = \nn \\
  & \frac{1}{8\pi^{2}\epsilon}\bigg\{
\big(18\lambda_{3Q^{\phi}}^{2}K_\phi^{-1}+ 12\lambda_{4Q^{\phi}}m_{h}^{2}\big)K_\phi^{-1} + \big(2\lambda^{2}_{Q^{\phi}2Q^{\theta}}K_\theta^{-1} + 2\lambda_{2Q^{\phi}2Q^{\theta}}m_{\theta}^{2}\big) n_{\rm GB} K_\theta^{-1} \nn \\
&+\sum_i \big( 6g_{2Q^{A^i}2Q^{\phi}}m_{A^i}^{2} + 6g_{2Q^{A^i}Q^{\phi}}^{2} + \frac{3}{4}K_\theta^{-1}\left(g_{Q^{A^i}\partial Q^{\phi}Q^{\theta^i}} + g_{Q^{A^i}\partial Q^{\theta^i} Q^{\phi}} \right)^{2}p^{2} -6  \mathcal{K}_{2Q^{\phi} 2 \partial Q^{A^i}} m_{A^i}^4\big)  K_A^{-1}\nn \\
  &- \big(2y_{Q^{\phi}}^{2}( p^{2} + 6m_{\psi}^{2} )K_{\psi}^{-1} + 8y_{2Q^{\phi}}m_{\psi}^{3}\big)N_c K_{\psi}^{-1} \bigg\}  - \Big[ \left( Z_{\partial Q^{\phi}} -1 \right)p^{2}K_\phi+ \left( Z_{\lambda 2Q^{\phi}} - 1 \right)2\lambda_{2Q^{\phi}} \Big].
\label{Pi_h_SM}
\end{align}
The Goldstone self-energy is generalized similarly.

\paragraph{Top quark self energy: $\Pi_{\psi_f}$.}
Consider first the Higgs/Goldstone contribution.  The Yukawa interaction is now of the form
\be
{\cal L}_y = - y (\bar t_L \phi_0^* t_R - \bar b_L \phi^- t_R + {\rm h.c.} )
\ee
with $\phi_0 = \phi + i \theta_1$ and $\phi^- =\theta_2-i \theta_3$.  The first term is the same as in the Abelian model which reproduces the first four terms in \cref{Pi_psi}.  The second term gives an additional contribution to $\bar \psi P_R \psi$, as it can give loops with $\phi^-$ and $b_L$ in the loop. This gives an additional contribution of the same form as the first line in \cref{Pi_psi}, but for $\phi,\theta \to \theta_1,\theta_2$ and $m_t \to m_b$ -- except from the $K_I^{-1}$-factors, these terms are the same as in the SM regime.  The first two terms on the 2nd line of \cref{Pi_psi}, which only arise in the large field regime, originate from the $\phi^0$-coupling and are the same as in the U(1) model.

For the gauge contribution, the last term on the 2nd line in \cref{Pi_psi}, to be proportional to the top mass, the gauge interactions have to be diagonal.  We thus include the $Z$, $\gamma$ and the QCD contribution.  We replace
\be
g^2 q_L q_R \to \sum_{i=\gamma,Z} q^i_L q^i_R + g_s^2 C_2(N_c)
\ee
where $q_{L,R}^i$ is the coupling of the left/right-handed top quark to $A^i$ (the gauge couplings are absorbed), and $g_s$ is the QCD coupling $C_2(N) = (N^2-1)/(2N)$. Except from the $K_A^{-1}$-factor these terms are the same the usual SM results.

Although the coefficients of the various terms gets modified in the full SM case, the order in $\delta$ of the various terms remains the same.  Thus also in the full SM we find for the current set-up that $Z_{y 2\psi} = Z_{2\psi_R} =Z_{2\psi_L} ={\cal O}(\delta^{-1}) $.

\paragraph{Gauge boson self energies: $\Pi_{A}$}

As noted before, in all loops containing gauge-Higgs interactions we can use a $U(1)$ model and sum over the three gauge couplings. This takes care of the first three lines in \cref{Pi_A}. To generalize the contribution from the fermion line we have to add the appropiate group factors. Finally, there are also new diagrams with gauge loops due to the non-abelian gauge interactions that can contribute to the wave-function normalization. Using the parameteric arguments in the paragraph below \cref{Pi_new}, it is easy to see that these contributions are subdominant.  Hence, just as for the fermion self-energy we conclude that for our set-up the SM generalization may change the coefficients and add new terms to the self-energy, but all terms can be neglected at leading order.

With the couplings given in \ref{A:couplings}, the self-energies in the large field limit in the full SM are for case B:
\begin{align}
\Pi^{\rm SM}_{Q^{\phi}} &= p^2 \delta\[ -(Z_{2Q^{\phi}}-1) + \mathcal{O}(\delta ^{-1})\] + \phi^2\[ -(Z_{\lambda 2Q^{\phi}}-1) \lambda + \frac1{8\pi^2 \eps} \(\sum_i \frac{3 g_i^4}{\alpha_0} -N_c y^4\)\]
\nn \\
\Pi^{\rm SM}_{Q^{\theta}} &=p^2 \delta\[ -(Z_{2Q^{\theta}}-1) + \mathcal{O}(\delta ^{-1})\] + 3\phi^2\[ -(Z_{\lambda 2Q^{\theta}}-1) \lambda + \frac1{8\pi^2 \eps} \(\sum_i \frac{3 g_i^4}{\alpha_0} - N_cy^4\)\]
\nn\\\
\Pi^{\rm SM}_\psi &= \slashed{p}\[ -P_L(Z_{2\psi_L}-1) -P_R(Z_{2\psi_R}-1) + \mathcal{O}(\delta ^{-1})\] + \phi\[ -(Z_{m 2\psi}-1) \frac{y}{\sqrt{2}} +  \mathcal{O}(\delta ^{-1})\]
\nn \\
  \Pi^{A,{\rm SM}}_{\mu\nu} &=(p^2 g_{\mu\nu} - p_\mu p_\nu)\delta \alpha_0 \[ -(Z_{2Q^A}-1) + \mathcal{O}(\delta ^{-1})\] + g_{\mu\nu} \phi^2 \delta \[ -(Z_{m^2_{A}}-1) g^2+  \mathcal{O}(\delta ^{-1})\].
                   \label{answer_abel}
\end{align}
Case A can be obainted by setting $g\to 0$, case C by setting $y \to 0$, and case D by setting $y,g \to 0$.

\bibliographystyle{utphys}
\bibliography{biblioMatching}

\providecommand{\href}[2]{#2}\begingroup\raggedright\begin{thebibliography}{10}

\bibitem{Akrami:2018odb}
{\bfseries Planck} Collaboration, Y.~Akrami {\em et~al.}, ``{Planck 2018
  results. X. Constraints on inflation},''
\href{http://arxiv.org/abs/1807.06211}{{\ttfamily arXiv:1807.06211
  [astro-ph.CO]}}.

\bibitem{Martin:2013tda}
J.~Martin, C.~Ringeval, and V.~Vennin, ``{Encyclopædia Inflationaris},''
  \href{http://dx.doi.org/10.1016/j.dark.2014.01.003}{{\em Phys. Dark Univ.}
  {\bfseries 5-6} (2014) 75--235},
\href{http://arxiv.org/abs/1303.3787}{{\ttfamily arXiv:1303.3787
  [astro-ph.CO]}}.

\bibitem{Copeland:1994vg}
E.~J. Copeland, A.~R. Liddle, D.~H. Lyth, E.~D. Stewart, and D.~Wands, ``{False
  vacuum inflation with Einstein gravity},''
  \href{http://dx.doi.org/10.1103/PhysRevD.49.6410}{{\em Phys. Rev.} {\bfseries
  D49} (1994) 6410--6433},
\href{http://arxiv.org/abs/astro-ph/9401011}{{\ttfamily arXiv:astro-ph/9401011
  [astro-ph]}}.

\bibitem{Baumann:2014nda}
D.~Baumann and L.~McAllister,
  \href{http://dx.doi.org/10.1017/CBO9781316105733}{{\em {Inflation and String
  Theory}}}.
\newblock Cambridge Monographs on Mathematical Physics. Cambridge University
  Press, 2015.
\newblock \href{http://arxiv.org/abs/1404.2601}{{\ttfamily arXiv:1404.2601
  [hep-th]}}.
\newblock
\url{http://www.cambridge.org/mw/academic/subjects/physics/theoretical-physics-and-mathematical-physics/inflation-and-string-theory?format=HB}.
\newblock

\bibitem{cliffnew}
C.~P. Burgess, S.~P. Patil, and M.~Trott, ``{On the Predictiveness of
  Single-Field Inflationary Models},''
  \href{http://dx.doi.org/10.1007/JHEP06(2014)010}{{\em JHEP} {\bfseries 06}
  (2014) 010},
\href{http://arxiv.org/abs/1402.1476}{{\ttfamily arXiv:1402.1476 [hep-ph]}}.

\bibitem{Fumagalli}
J.~Fumagalli and M.~Postma, ``{UV (in)sensitivity of Higgs inflation},''
  \href{http://dx.doi.org/10.1007/JHEP05(2016)049}{{\em JHEP} {\bfseries 05}
  (2016) 049},
\href{http://arxiv.org/abs/1602.07234}{{\ttfamily arXiv:1602.07234 [hep-ph]}}.

\bibitem{Germani:2010gm}
C.~Germani and A.~Kehagias, ``{New Model of Inflation with Non-minimal
  Derivative Coupling of Standard Model Higgs Boson to Gravity},''
  \href{http://dx.doi.org/10.1103/PhysRevLett.105.011302}{{\em Phys. Rev.
  Lett.} {\bfseries 105} (2010) 011302},
\href{http://arxiv.org/abs/1003.2635}{{\ttfamily arXiv:1003.2635 [hep-ph]}}.

\bibitem{Germani:2011cv}
C.~Germani, ``{Spontaneous localization on a brane via a gravitational
  mechanism},'' \href{http://dx.doi.org/10.1103/PhysRevD.85.055025}{{\em Phys.
  Rev.} {\bfseries D85} (2012) 055025},
\href{http://arxiv.org/abs/1109.3718}{{\ttfamily arXiv:1109.3718 [hep-ph]}}.

\bibitem{disformal}
S.~Di~Vita and C.~Germani, ``{Electroweak vacuum stability and inflation via
  nonminimal derivative couplings to gravity},''
  \href{http://dx.doi.org/10.1103/PhysRevD.93.045005}{{\em Phys. Rev.}
  {\bfseries D93} no.~4, (2016) 045005},
\href{http://arxiv.org/abs/1508.04777}{{\ttfamily arXiv:1508.04777 [hep-ph]}}.

\bibitem{trott}
E.~E. Jenkins, A.~V. Manohar, and M.~Trott, ``{Renormalization Group Evolution
  of the Standard Model Dimension Six Operators I: Formalism and lambda
  Dependence},'' \href{http://dx.doi.org/10.1007/JHEP10(2013)087}{{\em JHEP}
  {\bfseries 10} (2013) 087},
\href{http://arxiv.org/abs/1308.2627}{{\ttfamily arXiv:1308.2627 [hep-ph]}}.

\bibitem{shap}
F.~Bezrukov, J.~Rubio, and M.~Shaposhnikov, ``{Living beyond the edge: Higgs
  inflation and vacuum metastability},''
  \href{http://dx.doi.org/10.1103/PhysRevD.92.083512}{{\em Phys. Rev.}
  {\bfseries D92} no.~8, (2015) 083512},
\href{http://arxiv.org/abs/1412.3811}{{\ttfamily arXiv:1412.3811 [hep-ph]}}.

\bibitem{critical1}
F.~Bezrukov and M.~Shaposhnikov, ``{Higgs inflation at the critical point},''
  \href{http://dx.doi.org/10.1016/j.physletb.2014.05.074}{{\em Phys. Lett.}
  {\bfseries B734} (2014) 249--254},
\href{http://arxiv.org/abs/1403.6078}{{\ttfamily arXiv:1403.6078 [hep-ph]}}.

\bibitem{Enckell:2016xse}
V.-M. Enckell, K.~Enqvist, and S.~Nurmi, ``{Observational signatures of Higgs
  inflation},'' \href{http://dx.doi.org/10.1088/1475-7516/2016/07/047}{{\em
  JCAP} {\bfseries 1607} no.~07, (2016) 047},
\href{http://arxiv.org/abs/1603.07572}{{\ttfamily arXiv:1603.07572
  [astro-ph.CO]}}.

\bibitem{Bezrukov:2017dyv}
F.~Bezrukov, M.~Pauly, and J.~Rubio, ``{On the robustness of the primordial
  power spectrum in renormalized Higgs inflation},''
\href{http://arxiv.org/abs/1706.05007}{{\ttfamily arXiv:1706.05007 [hep-ph]}}.

\bibitem{Fumagalli:2017cdo}
J.~Fumagalli, S.~Mooij, and M.~Postma, ``{Unitarity and predictiveness in new
  Higgs inflation},'' \href{http://dx.doi.org/10.1007/JHEP03(2018)038}{{\em
  JHEP} {\bfseries 03} (2018) 038},
\href{http://arxiv.org/abs/1711.08761}{{\ttfamily arXiv:1711.08761 [hep-ph]}}.

\bibitem{bezrukov1}
F.~L. Bezrukov and M.~Shaposhnikov, ``{The Standard Model Higgs boson as the
  inflaton},'' \href{http://dx.doi.org/10.1016/j.physletb.2007.11.072}{{\em
  Phys. Lett.} {\bfseries B659} (2008) 703--706},
\href{http://arxiv.org/abs/0710.3755}{{\ttfamily arXiv:0710.3755 [hep-th]}}.

\bibitem{Kamada:2012se}
K.~Kamada, T.~Kobayashi, T.~Takahashi, M.~Yamaguchi, and J.~Yokoyama,
  ``{Generalized Higgs inflation},''
  \href{http://dx.doi.org/10.1103/PhysRevD.86.023504}{{\em Phys. Rev.}
  {\bfseries D86} (2012) 023504},
\href{http://arxiv.org/abs/1203.4059}{{\ttfamily arXiv:1203.4059 [hep-ph]}}.

\bibitem{Rubio:2018ogq}
J.~Rubio, ``{Higgs inflation},''
  \href{http://dx.doi.org/10.3389/fspas.2018.00050}{{\em Front. Astron. Space
  Sci.} {\bfseries 5} (2019) 50},
  \href{http://arxiv.org/abs/1807.02376}{{\ttfamily arXiv:1807.02376
  [hep-ph]}}.

\bibitem{bezrukov_loop}
F.~Bezrukov and M.~Shaposhnikov, ``{Standard Model Higgs boson mass from
  inflation: Two loop analysis},''
  \href{http://dx.doi.org/10.1088/1126-6708/2009/07/089}{{\em JHEP} {\bfseries
  07} (2009) 089},
\href{http://arxiv.org/abs/0904.1537}{{\ttfamily arXiv:0904.1537 [hep-ph]}}.

\bibitem{bezrukov4}
F.~Bezrukov, A.~Magnin, M.~Shaposhnikov, and S.~Sibiryakov, ``{Higgs inflation:
  consistency and generalisations},''
  \href{http://dx.doi.org/10.1007/JHEP01(2011)016}{{\em JHEP} {\bfseries 01}
  (2011) 016},
\href{http://arxiv.org/abs/1008.5157}{{\ttfamily arXiv:1008.5157 [hep-ph]}}.

\bibitem{wilczek}
A.~De~Simone, M.~P. Hertzberg, and F.~Wilczek, ``{Running Inflation in the
  Standard Model},''
  \href{http://dx.doi.org/10.1016/j.physletb.2009.05.054}{{\em Phys. Lett.}
  {\bfseries B678} (2009) 1--8},
\href{http://arxiv.org/abs/0812.4946}{{\ttfamily arXiv:0812.4946 [hep-ph]}}.

\bibitem{barvinsky}
A.~O. Barvinsky, A.~{\relax Yu}. Kamenshchik, and A.~A. Starobinsky,
  ``{Inflation scenario via the Standard Model Higgs boson and LHC},''
  \href{http://dx.doi.org/10.1088/1475-7516/2008/11/021}{{\em JCAP} {\bfseries
  0811} (2008) 021},
\href{http://arxiv.org/abs/0809.2104}{{\ttfamily arXiv:0809.2104 [hep-ph]}}.

\bibitem{barvinsky2}
A.~O. Barvinsky, A.~{\relax Yu}. Kamenshchik, C.~Kiefer, A.~A. Starobinsky, and
  C.~Steinwachs, ``{Asymptotic freedom in inflationary cosmology with a
  non-minimally coupled Higgs field},''
  \href{http://dx.doi.org/10.1088/1475-7516/2009/12/003}{{\em JCAP} {\bfseries
  0912} (2009) 003},
\href{http://arxiv.org/abs/0904.1698}{{\ttfamily arXiv:0904.1698 [hep-ph]}}.

\bibitem{barvinsky3}
A.~O. Barvinsky, A.~{\relax Yu}. Kamenshchik, C.~Kiefer, A.~A. Starobinsky, and
  C.~F. Steinwachs, ``{Higgs boson, renormalization group, and naturalness in
  cosmology},'' \href{http://dx.doi.org/10.1140/epjc/s10052-012-2219-3}{{\em
  Eur. Phys. J.} {\bfseries C72} (2012) 2219},
\href{http://arxiv.org/abs/0910.1041}{{\ttfamily arXiv:0910.1041 [hep-ph]}}.

\bibitem{damien}
D.~P. George, S.~Mooij, and M.~Postma, ``{Quantum corrections in Higgs
  inflation: the real scalar case},''
  \href{http://dx.doi.org/10.1088/1475-7516/2014/02/024}{{\em JCAP} {\bfseries
  1402} (2014) 024},
\href{http://arxiv.org/abs/1310.2157}{{\ttfamily arXiv:1310.2157 [hep-th]}}.

\bibitem{damien2}
D.~P. George, S.~Mooij, and M.~Postma, ``{Quantum corrections in Higgs
  inflation: the Standard Model case},''
  \href{http://dx.doi.org/10.1088/1475-7516/2016/04/006}{{\em JCAP} {\bfseries
  1604} no.~04, (2016) 006},
\href{http://arxiv.org/abs/1508.04660}{{\ttfamily arXiv:1508.04660 [hep-th]}}.

\bibitem{Hertzberg2}
M.~P. Hertzberg, ``{Can Inflation be Connected to Low Energy Particle
  Physics?},'' \href{http://dx.doi.org/10.1088/1475-7516/2012/08/008}{{\em
  JCAP} {\bfseries 1208} (2012) 008},
\href{http://arxiv.org/abs/1110.5650}{{\ttfamily arXiv:1110.5650 [hep-ph]}}.

\bibitem{mirage}
J.~L.~F. Barbon, J.~A. Casas, J.~Elias-Miro, and J.~R. Espinosa, ``{Higgs
  Inflation as a Mirage},''
  \href{http://dx.doi.org/10.1007/JHEP09(2015)027}{{\em JHEP} {\bfseries 09}
  (2015) 027},
\href{http://arxiv.org/abs/1501.02231}{{\ttfamily arXiv:1501.02231 [hep-ph]}}.

\bibitem{Fumagalli:2016sof}
J.~Fumagalli, ``{Renormalization Group independence of Cosmological
  Attractors},'' \href{http://dx.doi.org/10.1016/j.physletb.2017.04.017}{{\em
  Phys. Lett.} {\bfseries B769} (2017) 451--459},
\href{http://arxiv.org/abs/1611.04997}{{\ttfamily arXiv:1611.04997 [hep-th]}}.

\bibitem{Bauer:2008zj}
F.~Bauer and D.~A. Demir, ``{Inflation with Non-Minimal Coupling: Metric versus
  Palatini Formulations},''
  \href{http://dx.doi.org/10.1016/j.physletb.2008.06.014}{{\em Phys. Lett.}
  {\bfseries B665} (2008) 222--226},
\href{http://arxiv.org/abs/0803.2664}{{\ttfamily arXiv:0803.2664 [hep-ph]}}.

\bibitem{Rasanen:2017ivk}
S.~Rasanen and P.~Wahlman, ``{Higgs inflation with loop corrections in the
  Palatini formulation},''
  \href{http://dx.doi.org/10.1088/1475-7516/2017/11/047}{{\em JCAP} {\bfseries
  1711} no.~11, (2017) 047},
\href{http://arxiv.org/abs/1709.07853}{{\ttfamily arXiv:1709.07853
  [astro-ph.CO]}}.

\bibitem{Enckell:2018kkc}
V.-M. Enckell, K.~Enqvist, S.~Rasanen, and E.~Tomberg, ``{Higgs inflation at
  the hilltop},'' \href{http://dx.doi.org/10.1088/1475-7516/2018/06/005}{{\em
  JCAP} {\bfseries 1806} no.~06, (2018) 005},
\href{http://arxiv.org/abs/1802.09299}{{\ttfamily arXiv:1802.09299
  [astro-ph.CO]}}.

\bibitem{Rasanen:2018ihz}
S.~Rasanen, ``{Higgs inflation in the Palatini formulation with kinetic terms
  for the metric},''
\href{http://arxiv.org/abs/1811.09514}{{\ttfamily arXiv:1811.09514 [gr-qc]}}.

\bibitem{Racioppi:2019jsp}
A.~Racioppi, ``{Non-Minimal (Self-)Running Inflation: Metric vs. Palatini
  Formulation},'' \href{http://arxiv.org/abs/1912.10038}{{\ttfamily
  arXiv:1912.10038 [hep-ph]}}.

\bibitem{Bauer:2010jg}
F.~Bauer and D.~A. Demir, ``{Higgs-Palatini Inflation and Unitarity},''
  \href{http://dx.doi.org/10.1016/j.physletb.2011.03.042}{{\em Phys. Lett.}
  {\bfseries B698} (2011) 425--429},
\href{http://arxiv.org/abs/1012.2900}{{\ttfamily arXiv:1012.2900 [hep-ph]}}.

\bibitem{germaniU}
A.~Escrivà and C.~Germani, ``{Beyond dimensional analysis: Higgs and new Higgs
  inflations do not violate unitarity},''
  \href{http://dx.doi.org/10.1103/PhysRevD.95.123526}{{\em Phys. Rev.}
  {\bfseries D95} no.~12, (2017) 123526},
\href{http://arxiv.org/abs/1612.06253}{{\ttfamily arXiv:1612.06253 [hep-ph]}}.

\bibitem{Galante:2014ifa}
M.~Galante, R.~Kallosh, A.~Linde, and D.~Roest, ``{Unity of Cosmological
  Inflation Attractors},''
  \href{http://dx.doi.org/10.1103/PhysRevLett.114.141302}{{\em Phys. Rev.
  Lett.} {\bfseries 114} no.~14, (2015) 141302},
\href{http://arxiv.org/abs/1412.3797}{{\ttfamily arXiv:1412.3797 [hep-th]}}.

\bibitem{Vilkovisky:1984st}
G.~A. Vilkovisky, ``{The Unique Effective Action in Quantum Field Theory},''
\href{http://dx.doi.org/10.1016/0550-3213(84)90228-1}{{\em Nucl. Phys.}
  {\bfseries B234} (1984) 125--137}.

\bibitem{Gong:2011uw}
J.-O. Gong and T.~Tanaka, ``{A covariant approach to general field space metric
  in multi-field inflation},''
  \href{http://dx.doi.org/10.1088/1475-7516/2012/02/E01,
  10.1088/1475-7516/2011/03/015}{{\em JCAP} {\bfseries 1103} (2011) 015},
  \href{http://arxiv.org/abs/1101.4809}{{\ttfamily arXiv:1101.4809
  [astro-ph.CO]}}.
[Erratum: JCAP1202,E01(2012)].

\bibitem{DeWitt:1965jb}
B.~S. DeWitt, ``{Dynamical theory of groups and fields},'' {\em Conf. Proc.}
  {\bfseries C630701} (1964) 585--820.
[Les Houches Lect. Notes13,585(1964)].

\bibitem{Sasaki:1995aw}
M.~Sasaki and E.~D. Stewart, ``{A General analytic formula for the spectral
  index of the density perturbations produced during inflation},''
  \href{http://dx.doi.org/10.1143/PTP.95.71}{{\em Prog. Theor. Phys.}
  {\bfseries 95} (1996) 71--78},
  \href{http://arxiv.org/abs/astro-ph/9507001}{{\ttfamily
  arXiv:astro-ph/9507001}}.

\bibitem{GrootNibbelink:2001qt}
S.~Groot~Nibbelink and B.~van Tent, ``{Scalar perturbations during multiple
  field slow-roll inflation},''
  \href{http://dx.doi.org/10.1088/0264-9381/19/4/302}{{\em Class. Quant. Grav.}
  {\bfseries 19} (2002) 613--640},
  \href{http://arxiv.org/abs/hep-ph/0107272}{{\ttfamily arXiv:hep-ph/0107272}}.

\bibitem{Achucarro:2010da}
A.~Achucarro, J.-O. Gong, S.~Hardeman, G.~A. Palma, and S.~P. Patil,
  ``{Features of heavy physics in the CMB power spectrum},''
  \href{http://dx.doi.org/10.1088/1475-7516/2011/01/030}{{\em JCAP} {\bfseries
  01} (2011) 030}, \href{http://arxiv.org/abs/1010.3693}{{\ttfamily
  arXiv:1010.3693 [hep-ph]}}.

\bibitem{Alonso:2015fsp}
R.~Alonso, E.~E. Jenkins, and A.~V. Manohar, ``{A Geometric Formulation of
  Higgs Effective Field Theory: Measuring the Curvature of Scalar Field
  Space},'' \href{http://dx.doi.org/10.1016/j.physletb.2016.01.041}{{\em Phys.
  Lett. B} {\bfseries 754} (2016) 335--342},
  \href{http://arxiv.org/abs/1511.00724}{{\ttfamily arXiv:1511.00724
  [hep-ph]}}.

\bibitem{Alonso:2016oah}
R.~Alonso, E.~E. Jenkins, and A.~V. Manohar, ``{Geometry of the Scalar
  Sector},'' \href{http://dx.doi.org/10.1007/JHEP08(2016)101}{{\em JHEP}
  {\bfseries 08} (2016) 101}, \href{http://arxiv.org/abs/1605.03602}{{\ttfamily
  arXiv:1605.03602 [hep-ph]}}.

\bibitem{Helset:2018fgq}
A.~Helset, M.~Paraskevas, and M.~Trott, ``{Gauge fixing the Standard Model
  Effective Field Theory},''
  \href{http://dx.doi.org/10.1103/PhysRevLett.120.251801}{{\em Phys. Rev.
  Lett.} {\bfseries 120} no.~25, (2018) 251801},
  \href{http://arxiv.org/abs/1803.08001}{{\ttfamily arXiv:1803.08001
  [hep-ph]}}.

\bibitem{Nagai:2019tgi}
R.~Nagai, M.~Tanabashi, K.~Tsumura, and Y.~Uchida, ``{Symmetry and geometry in
  a generalized Higgs effective field theory: Finiteness of oblique corrections
  versus perturbative unitarity},''
  \href{http://dx.doi.org/10.1103/PhysRevD.100.075020}{{\em Phys. Rev. D}
  {\bfseries 100} no.~7, (2019) 075020},
  \href{http://arxiv.org/abs/1904.07618}{{\ttfamily arXiv:1904.07618
  [hep-ph]}}.

\bibitem{Helset:2020yio}
A.~Helset, A.~Martin, and M.~Trott, ``{The Geometric Standard Model Effective
  Field Theory},'' \href{http://dx.doi.org/10.1007/JHEP03(2020)163}{{\em JHEP}
  {\bfseries 03} (2020) 163}, \href{http://arxiv.org/abs/2001.01453}{{\ttfamily
  arXiv:2001.01453 [hep-ph]}}.

\bibitem{Alonso:2019mok}
R.~Alonso, ``{Covariant derivative expansion for the renormalization of
  gravity},'' \href{http://arxiv.org/abs/1912.09671}{{\ttfamily
  arXiv:1912.09671 [hep-ph]}}.

\bibitem{Falls:2018olk}
K.~Falls and M.~Herrero-Valea, ``{Frame (In)equivalence in Quantum Field Theory
  and Cosmology},''
  \href{http://dx.doi.org/10.1140/epjc/s10052-019-7070-3}{{\em Eur. Phys. J. C}
  {\bfseries 79} no.~7, (2019) 595},
  \href{http://arxiv.org/abs/1812.08187}{{\ttfamily arXiv:1812.08187
  [hep-th]}}.

\bibitem{Finn:2019aip}
K.~Finn, S.~Karamitsos, and A.~Pilaftsis, ``{Grand Covariance in Quantum
  Gravity},'' \href{http://arxiv.org/abs/1910.06661}{{\ttfamily
  arXiv:1910.06661 [hep-th]}}.

\bibitem{Fradkin:1983nw}
E.~S. Fradkin and A.~A. Tseytlin, ``{On the New Definition of Off-shell
  Effective Action},''
\href{http://dx.doi.org/10.1016/0550-3213(84)90075-0}{{\em Nucl. Phys.}
  {\bfseries B234} (1984) 509--523}.

\bibitem{DeWitt:1967ub}
B.~S. DeWitt, ``{Quantum Theory of Gravity. 2. The Manifestly Covariant
  Theory},'' \href{http://dx.doi.org/10.1103/PhysRev.162.1195}{{\em Phys. Rev.}
  {\bfseries 162} (1967) 1195--1239}.
[,298(1967)].

\bibitem{RGE}
D.~Buttazzo, G.~Degrassi, P.~P. Giardino, G.~F. Giudice, F.~Sala, A.~Salvio,
  and A.~Strumia, ``{Investigating the near-criticality of the Higgs boson},''
  \href{http://dx.doi.org/10.1007/JHEP12(2013)089}{{\em JHEP} {\bfseries 12}
  (2013) 089},
\href{http://arxiv.org/abs/1307.3536}{{\ttfamily arXiv:1307.3536 [hep-ph]}}.

\bibitem{Hoang:2014oea}
A.~H. Hoang, ``{The Top Mass: Interpretation and Theoretical Uncertainties},''
  in {\em Proceedings, 7th International Workshop on Top Quark Physics
  (TOP2014): Cannes, France, September 28-October 3, 2014}.
\newblock 2014.
\newblock
\href{http://arxiv.org/abs/1412.3649}{{\ttfamily arXiv:1412.3649 [hep-ph]}}.
\newblock

\bibitem{Nason:2017cxd}
P.~Nason, \href{http://dx.doi.org/10.1142/9789813238053_0008}{``{The Top Mass
  in Hadronic Collisions},''} in {\em From My Vast Repertoire ...: Guido
  Altarelli's Legacy}, A.~Levy, S.~Forte, and G.~Ridolfi, eds., pp.~123--151.
\newblock 2019.
\newblock
\href{http://arxiv.org/abs/1712.02796}{{\ttfamily arXiv:1712.02796 [hep-ph]}}.
\newblock

\bibitem{Tanabashi:2018oca}
{\bfseries Particle Data Group} Collaboration, M.~Tanabashi {\em et~al.},
  ``{Review of Particle Physics},''
\href{http://dx.doi.org/10.1103/PhysRevD.98.030001}{{\em Phys. Rev.} {\bfseries
  D98} no.~3, (2018) 030001}.

\bibitem{Luo:2002ey}
M.-x. Luo and Y.~Xiao, ``{Two loop renormalization group equations in the
  standard model},''
  \href{http://dx.doi.org/10.1103/PhysRevLett.90.011601}{{\em Phys. Rev. Lett.}
  {\bfseries 90} (2003) 011601},
\href{http://arxiv.org/abs/hep-ph/0207271}{{\ttfamily arXiv:hep-ph/0207271
  [hep-ph]}}.

\bibitem{Urbano:2019ohp}
A.~Urbano, ``{Inflation without gauge redundancy},''
  \href{http://dx.doi.org/10.1088/1475-7516/2020/04/040}{{\em JCAP} {\bfseries
  04} (2020) 040}, \href{http://arxiv.org/abs/2001.05480}{{\ttfamily
  arXiv:2001.05480 [hep-th]}}.

\bibitem{Planck}
{\bfseries Planck} Collaboration, P.~A.~R. Ade {\em et~al.}, ``{Planck 2015
  results. XX. Constraints on inflation},''
\href{http://arxiv.org/abs/1502.02114}{{\ttfamily arXiv:1502.02114
  [astro-ph.CO]}}.

\bibitem{Shaposhnikov:2020fdv}
M.~Shaposhnikov, A.~Shkerin, and S.~Zell, ``{Quantum Effects in Palatini Higgs
  Inflation},'' \href{http://arxiv.org/abs/2002.07105}{{\ttfamily
  arXiv:2002.07105 [hep-ph]}}.

\end{thebibliography}\endgroup

\end{document}